\documentclass[12pt]{iopart}
\usepackage{iopams}
\usepackage{graphicx}
\usepackage{epsfig}
\usepackage{hyperref}

\begin{document}

\title[Complex Solitary Wave Dynamics]{\bf{Complex Solitary Wave Dynamics, Pattern Formation, and Chaos in the Gain-Loss Nonlinear Schr\"odinger Equation}}

\author{Justin Q. Anderson$^1$, Rachel A. Ryan$^1$, Mingzhong Wu$^2$ and Lincoln D Carr$^1$}
\address{$^1$ Department of Physics, Colorado School of Mines, Golden, CO 80401, USA}
\address{$^2$ Department of Physics, Colorado State University, Fort Collins, CO 80523, USA}
\ead{jusander@mines.edu}

\date{\today}

\begin{abstract}
A numerical exploration of a gain-loss nonlinear Schr\"odinger equation was carried out utilizing over 180000 core hours to conduct more than 10000 unique simulations in an effort to characterize the model's six dimensional parameter space. The study treated the problem in full generality, spanning a minimum of eight orders of magnitude for each of three linear and nonlinear gain terms and five orders of magnitude for higher order nonlinearities.  The gain-loss nonlinear Schr\"odinger equation is of interest as a model for spin wave envelopes in magnetic thin film active feedback rings and analogous driven damped nonlinear physical systems.  Bright soliton trains were spontaneously driven out of equilibrium and behaviors stable for tens of thousands of round trip times were numerically identified. Nine distinct complex dynamical behaviors with lifetimes on the order of ms were isolated as part of six identified solution classes. Numerically located dynamical behaviors include: (1) Low dimensional chaotic modulations of bright soliton trains; (2) spatially symmetric/asymmetric interactions of solitary wave peaks; (3) dynamical pattern formation and recurrence; (4) steady state solutions and (5) intermittency.  Simulations exhibiting chaotically modulating bright soliton trains were found to qualitatively match previous experimental observations. Ten new dynamical behaviors, eight demonstrating long lifetimes, are predicted to be observable in future experiments.
\end{abstract}

\maketitle

\section{\label{sec:intro}Introduction}

Related forms of the nonlinear Schr\"odinger equation are used to explore nonlinear phenomena in many distinct physical systems: Ginzburg-Landau equations describe the envelope evolution of mode-locked lasers, and superconductivity~\cite{Arnason:2002}; the cubic nonlinear Schr\"odinger equation treats deep water waves~\cite{Sulem:1999} and the dynamics of spin-wave envelopes in magnetic thin films~\cite{Kalinikos:1986,Slavin:1987}; a driven damped nonlinear Schr\"odinger equation models exciton-polariton and magnon Bose-Einstein condensates (BECs)~\cite{Carusotto:2013}; and the Gross-Pitaevskii equation models the mean field of atomic and molecular BECs~\cite{Pitaevskii:2003,Carr:2009}. With the increasing supply of cheap computing power these systems have become the subject of extensive and sometimes rigorous numerical study.

Magnetic thin films have demonstrated potential as a versatile toy system for experiments on fundamental nonlinear dynamics~\cite{Wu:2010}.  Over the past two decades yttrium-iron-garnet ($\mathrm{Y}_{3}\mathrm{Fe}_{5}\mathrm{O}_{12}$, YIG) magnetic thin films have been fruitfully studied by numerous experimental groups and have demonstrated a rich variety of nonlinear phenomena. These include bright and dark envelope solitons~\cite{Slavin:1994,Scott:2005,Kostylev:2005,Kalinikos:1999,Koshikov:1996,Kalinikos:2000,Kalinikos:1991,Kalinikos:1990,Kalinikos:1990-2,Chen:1994,Kalinikos:1998,Kalinikos:1991,Kalinikos:1992,Benner:2000,Slavin:2003}, soliton trains~\cite{Wu:2004,Kalinikos:2002,Kalinikos:1998-2}, m\"obius solitons~\cite{Demokritov:2003}, Fermi-Pasta-Ulam and spatial recurrence~\cite{Wu:2007,Scott:2003}, soliton fractals~\cite{Wu:2006-2}, random solitons~\cite{Wu:2006}, chaotic spin waves~\cite{Wu:2005,Hagerstrom:2009,Kondrashov:2008}, multiple solitons~\cite{Wu:2004-2}, and chaotic solitons~\cite{Wang:2011,Ustinov:2011}.

A majority of these phenomena were observed on active feedback rings; such feedback structures are ubiquitous within science and physics in general.  Rings in particular are commonly used to study wave dynamics when one seeks a quantized wavenumber, periodic pumping, self-generation, or other resonant phenomena. Active rings, so called for the presence of periodic linear amplification, allow for the direct compensation of the major loss mechanisms present within a system.  This permits one to drive the system into quasi-conservative regimes, enabling the observation of dynamics on scales of several to tens of thousands of round trip times.  Dynamics with lifetimes of this order would otherwise be prohibited by the presence of dissipation. 

Yet, within the context of spin waves in magnetic thin films, little work has been carried out to develop an adequate theory for describing the rich range of behaviors evident within these recent experimental works. The integrable cubic nonlinear Schr\"odinger equation~(NLS), while successful in quantitatively describing both dark and bright soliton trains, is unable to reproduce more complex phenomena such as the chaotic oscillation of soliton envelopes. However, there has been significant effort within the mode-locked laser community to study analogous driven and damped systems. Works on dissipation terms and saturation~\cite{Ablowitz:2001,Ablowitz:2008,Ablowitz:2008-2,Ablowitz:2008-3}, the study of dissipative solitons dynamics~\cite{Akhmediev:2001,Akhmediev:2005,Akhmediev:2007,Akhmediev:2008,Akhmediev:2009} and other numerical studies of the cubic quintic complex Ginzburg-Landau equation~\cite{Tsoy:2005,Zhuravlev:2004} are highly relevant to the development of a driven damped model for spin waves in magnetic thin films. These works explore the dynamics of solitary waves and their associated wave equations under the influences of gain and loss. For example, the dynamics of near steady-state dissipative solitons have been considered in detail; such studies include rigorous mappings of stable and unstable regions of parameter space~\cite{Akhmediev:2007,Akhmediev:2008,Akhmediev:2009}. Similarly, initial transient behaviors have been the subject of significant research efforts by the mode-locked laser community. Transients are of interest for potential applications in signal processing and communication. To date there have been no efforts towards the characterization of long lifetime $(>1~\mathrm{ms})$ dynamics of soliton trains driven from equilibrium within active feedback rings. The work presented here demonstrates such an effort for a generalized nonlinear Schr\"odinger equation with a focus on applications to nonlinear  spin-waves propagating in an active feedback ring.  

Our paper is outlined as follows.~\Sref{sec:meth} introduces the model to be studied along with the associated operating limits; here the methodology and scope of the simulations are explicitly defined.  Experimental contexts for the work are also considered in~\sref{sec:meth}. Results in the form of eleven unique complex dynamical behaviors are presented and categorized in sections~\ref{sec:chaos}-\ref{sec:intermittent}.~\Sref{sec:chaos} contains simulations of chaotically modulating soliton trains. Spatially symmetric and asymmetric solitary wave interactions are presented in \sref{sec:interactions}. Four examples of dynamical pattern formation are given in~\sref{sec:DPF}. Two cases of steady state evolution are reported in~\sref{sec:steady_state}. Intermittent solutions are discussed in~\sref{sec:intermittent}. Finally, a discussion of numerical convergence and solution robustness is given in~\sref{appendix}. The work is summarized in~\sref{sec:conc}. Animations of each dynamical behavior discussed in sections~\ref{sec:chaos}-\ref{sec:intermittent} are available at \url{http://mines.edu/~jusander/GLNLS_dynamics/}.

\section{\label{sec:meth}Model Overview}
Motivated by the works discussed in~\sref{sec:intro}, we propose a generalized governing equation for spin waves in magnetic thin film active feedback rings: the \textit{gain loss nonlinear Schr\"odinger equation} (GLNLS),

\begin{equation}\label{eqn:model:GLNLS}
i\frac{\partial u}{\partial t}=\left[-\frac{D}{2}\frac{\partial^2}{\partial x^2} + iL + (N+iC)\vert u\vert^2 +(S+iQ)\vert u\vert^4\right]u
\end{equation}
where $u=u(x,t)$ is a dimensionless complex magnetization amplitude defined as $|u(x,t)|^2=\mathbf{m}(x,t)^2/2\mathbf{M_{\mathrm{s}}}^2$; here $\mathbf{m}(x,t)$ is the dynamic magnetization while $\mathbf{M_{\mathrm{s}}}^2$ is the saturation magnetization; $D$ is the dispersion coefficient; $N$ and $S$ are the cubic and quintic nonlinearity coefficients, respectively; $t$ is the `temporal' evolution coordinate; $x$ is the `spatial' coordinate of propagation boosted to the group velocity of the envelope; and $L$, $C$, and $Q$ are the linear, cubic, and quintic gains (losses) if positive (negative).  All parameters are taken to be real as the complex nature of the coefficients is explicitly accounted for in~\eref{eqn:model:GLNLS}. The local intensity of the magnetization amplitude is given by $|u(x,t)|^2$. The norm and energy at a given time, $t$, are defined as
\begin{equation}\label{eqn:model:norm}
\| u(t)\|_{2}=\int_{0}^{L}\!\! \rmd x\,\, |u(x,t)|^2,
\end{equation}
and 
\begin{equation}\label{eqn:model:energy}
 E(t)=\int^{L}_{0}\!\! \rmd x\, \left[\frac{D}{2} \left|\frac{\partial u(x,t)}{\partial x}\right|^2 + (N+S|u(x,t)|^2)|u(x,t)|^2\right],
\end{equation}
respectively, where the integrals are taken over the length or circumference, $2\pi R$, of the feedback ring.  All norms, intensities and energies given within figures and animations are scaled by $||u(0)||_{2}$, $\mathrm{max}\left[|u(x,0)|^{2}\right]$ and $\mathrm{abs}\left[E(0)\right]$ respectively where $t=0$ corresponds to the initial condition used during numerical simulation. Numerical values given within the text are not scaled. The specific choice of initial condition is discussed later in this section.

The gain and loss present within the GLNLS may be viewed as an expansion of saturable loss expressions studied separately by Ablowitz and Akhmediev~\cite{Ablowitz:2008,Ablowitz:2008-2,Akhmediev:2007,Akhmediev:2008}.  The higher order nonlinearity, $S$, may be used either as a saturation of cubic nonlinearity or an additional self-steepening; both cases are studied in the literature~\cite{Sulem:1999}. The GLNLS omits other terms commonly included in cubic quintic complex Ginzburg-Landau equations such as spectral filtering, periodic pumping, and integral mean terms, as they are not needed in this physical context. We are likewise compelled by Occam's Razor to choose the simplest possible model which nevertheless reproduces measurements in magnetic thin film active feedback rings.  NLS-like equations may be derived in magnetic thin films by use of a slowly varying envelope approximation~\cite{Stancil:2009}, more rigorously through conservation considerations and a Hamiltonian formalism~\cite{Slavin:1994,Krivosik:2012}, or directly from Maxwell's equations using multi-scale methods~\cite{Leblond:2001}.   
\begin{figure}[ht]
\centering
\includegraphics[width=.5\textwidth]{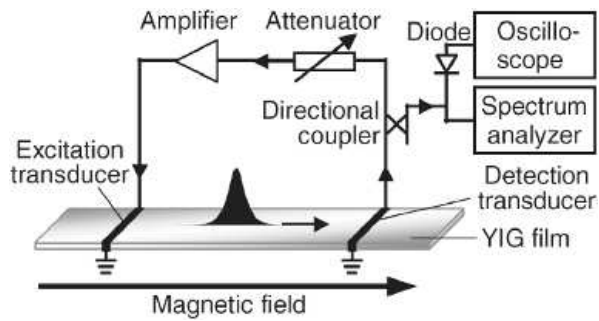}
\caption{\label{fig:feedback_ring}Diagram of active feedback ring experimental apparatus.  Reprinted with permission~\cite{Wu:2006-2}.}
\end{figure}
The operating limits of the GLNLS are motivated principally by experimental work on the excitation of chaotic solitons in YIG strip-based active feedback rings~\cite{Wang:2011}.  A block diagram of the active feedback ring experiment is shown in~\fref{fig:feedback_ring}. The ring is comprised of a nonlinear propagation medium, in this case a magnetically saturated crystalline YIG thin film, connected via two transducers to an electronic feedback loop. The electronics loop is constructed of a directional coupler, allowing real time observation at an oscilloscope and/or spectrum analyzer, and an amplifier/attenuator pair for real time adjustment of ring gain. The GLNLS demonstrated qualitative agreement with the low dimensional chaotic modulation of a bright soliton train, as will be discussed in~\sref{sec:chaos}. Detailed experimental results are discussed in Wang \etal.~\cite{Wang:2011}. These experiments indicate that nonlinearity and dispersion are the dominant sources of envelope shaping for chaotic spin wave solitons and that the losses present in the ring are fully compensated for by the amplifier. This imposed two constraints on modeling. (1) The coefficients $N$ and $D$ must be orders of magnitude larger than $L$, $C$, and $Q$. (2) The linear amplifier must compensate both the linear and nonlinear losses present in the film, requiring a net averaged (over many round trips) linear gain, $L>0$. Likewise, the dissipative terms represent the net gain and loss processes occurring in the ring averaged over several round trip times. One expects the use of this approximation, and therefore the model, to be valid when the time scale of envelope modulation is much greater than the soliton round trip time. All simulations were performed using adaptive time step Cash-Karp Runge-Kutta for temporal evolution and pseudospectral techniques for spatial propagation~\cite{Snyder:2008,NR}. Periodic boundary conditions modeled propagation around a ring. 

Parameter space explorations were explicitly chosen to encompass the GLNLS operating regime for magnetic thin film systems, while extending into other limits that could be of interest to other systems where the GLNLS is a useful model. Along with the previously mentioned restriction on the sign of $L$ only cases with cubic losses, $C\le0$, were considered. Both instances of saturating\footnote{A typical expression for saturable gain is given by $iS_{\mathrm{g}}\left(1+\frac{|u(x,t)|^2}{I_{\mathrm{s}}}\right)^{-1}$ where $I_{\mathrm{s}}$ and $S_{\mathrm{g}}$ are control parameters. Expanding denominator to third order yields $iS_{\mathrm{g}}\left(1-|u(x,t)|^2I_{\mathrm{s}}^{-1}+|u(x,t)|^4I_{\mathrm{s}}^{-2}+...\right)$, hence positive quintic gain being named \it{saturating}.} quintic gains, $Q\ge0$, and supplemental quintic losses, $Q\le0$, were studied. No sign restrictions were placed on quintic nonlinearity, $S$. The terms were explored in a decadal fashion across the GLNLS scaled values listed here
\begin{itemize}
\item $L=10^{n},~~n\in\left\{0,-1,-2,-3,-4,-5,-6,-7\right\}$,
\item $C\in\left\{0,-10^{n}\right\},~~n\in\left\{0,-1,-2,-3,-4,-5,-6\right\}$,
\item $Q\in\left\{0,\pm 10^{n}\right\},~~n\in\left\{0,-1,-2,-3,-4,-5,-6,-7\right\}$,
\item $S\in\left\{0,\pm10^{-1},\pm10^{-2}\right\}$,
\end{itemize}
for a total of eight possible values of $L$ and $C$, five choices for $S$, and 17 unique choices for $Q$. Ignoring cases with solely gains present we performed 5,470 unique simulations. An additional 1,530 simulations were undertaken with random parameters. The value for any single parameter in these simulations was generated by multiplying a pseudo random number between zero and one, from the uniform distribution, by an order of magnitude and sign chosen at random, again with uniform weight, from a parameter's allowed values, as defined above. To avoid ambiguity all statements in this paper concerning the relative size of GLNLS parameters refer to the order of magnitude and not the sign.

Simulations began as a bright soliton initial condition obtained via imaginary time relaxation~\cite{Kosloff:1986}, the ground state solution to the GLNLS with $S$, $L$, $C$, and $Q$ set as zero, with $|u(x,t)|^2<1$. Experimentally this corresponds to a stable bright soliton circling within a YIG strip-based active feedback ring, a solution analogous to a soliton train. Physical GLNLS parameter values are obtained by fitting this initial condition to experimentally observed bright soliton train conditions. This choice of units also fixes the ratio of $N/D$ used in simulations, while the amplitude of $N$ and $D$ dictate the simulation timescale. We assumed that the dimensionless spin-wave intensity is directly proportional to the spin-wave power, $|u(x,t)|^2\propto P_{\mathrm{out}}$, since experimental measurements of voltage are taken across a diode with quadratic behavior and are generally taken to be proportional to power. Values typical for a chaotic soliton experiment are $T=165~\mathrm{ns}$, the round trip time; $d=0.55~\mathrm{cm}$, the transducer separation; $T_{\mathrm{e}}=10~\mathrm{ns}$, the electronic loop propagation time; $V_{\mathrm{g}}=d/(T-T_{\mathrm{e}})=3.5\times10^6~\mathrm{cm/s}$, the group velocity; $N=-9.24\times10^9~\mathrm{rad/s}$, the cubic nonlinearity;  and $D=510 ~\mathrm{cm^2/s}$, the dispersion. Using these parameters one finds $\left[ t\right]\approx 25~\mathrm{ns}$ where $t$ is the scaled temporal unit used in simulations. This relation may be used to immediately transform code values for $L$, $C$, $Q$ and $S$, which share units of inverse time, to physical values. For example the largest studied linear gain is $L=t^{-1}\approx0.05\,\mathrm{ns}^{-1}$ which matches the order of experimentally approximated linear losses for magnetostatic backward volume spin waves in YIG thin films~\cite{Scott:2004}.  Experimentally a time series is recorded at the detection transducer with the full waveform being captured once a round trip after the signal has propagated a length $d$ between the transducers and passed through the electronics loop. The length of the ring, $\ell$, is taken to be the transducer separation, $d$, as the propagation delay is orders of magnitude smaller than the round trip time, $T_{\mathrm{e}}<T$. Simulations explicitly model the entire feedback loop at the group velocity of the waveform.  A time series may be reconstructed from numerical data by concatenating the simulated waveform after a temporal evolution of $T$ or a spatial evolution of $d=\ell$. In this work we adopt the former convention to ease the direct comparison of simulations to the power vs. time data often observed experimentally for spin waves in magnetic thin films. Such a reconstructed time series is labeled $u_{\mathrm{ts}}(t)$ throughout the paper. A time series of solitary wave peak intensity at successive round trips is useful in studying modulating single solitary wave trains and is defined by
\begin{equation}\label{eqn:upeak}
 |u_{\mathrm{peak}}(t)|^2=|\max\left[ u(x,nT) \right]|^2,\,\,n=0,1,...,N_{\mathrm{RT}},
\end{equation}
where $T$ is the round trip time and $N_{\mathrm{RT}}$ is the total number of round trips.

Over 180000 core hours were utilized to conduct more than 10000 unique simulations and convergence studies. An initial study of 3500 simulations was undertaken to explore the extent of transient effects and the numerical convergence behaviors of the GLNLS. A summary and analysis of the subsequent 7000 simulations, corresponding to over 3~TB of data, are presented in sections~\ref{sec:chaos}-\ref{sec:intermittent}. Approximately 1500 simulations were evaluated in detail; the remaining simulations were spot checked for consistency. Dozens of complex dynamical behaviors were identified during the course of simulation. We call this system complex because it displays a rich variety of dynamical behaviors, including chaos, robust emergent solitary-wave features, and generally multiple scales in both space and time. Solution types were divided into three stability cases, with each case corresponding to roughly $30\%$ of observed dynamics. The three cases are stable, intermittent and unstable. Temporally stable solutions demonstrated substantial observable lifetimes, greater than $1~\mathrm{ms}$ or 7000+ round trips, and robustness to variations in initial conditions of at least $10\%$. Evolution was found to be least sensitive to changes in $S$ and $Q$ and most sensitive to perturbations in $L$. In general the effect of changes in initial conditions tended to degrade the lifetime of dynamical behaviors and push solutions towards the intermittent case.  Nine temporally stable distinct dynamics and two separate cases of intermittency are discussed below. 

\section{\label{sec:chaos}Chaotic Modulation}

The chaotic modulation of stable solitary wave trains was observed for solutions containing strongly saturated cubic nonlinearity, $S\geq10^{-2}$, and the lowest studied ring gains, $L=10^{-7}$, with matching orders of cubic and/or quintic losses. A single bright soliton is observed to circulate within an active feedback ring while exhibiting complex modulations in peak intensity. Low ring losses are anticipated for this solution type, as experimentally observed chaotically modulating soliton trains have lifetimes measured in seconds. The presence of a single stable bright soliton suggests that nonlinearity and dispersion are the dominate forces in peak shaping. These are two conditions used during the derivation of the GLNLS,~\eref{eqn:model:GLNLS}, as discussed previously in~\sref{sec:meth}.

The chaotic nature of measured time series was verified by using standard phase space reconstruction techniques available in the open source Nonlinear Time Series (TISEAN) package to arrive at a stable correlation dimension, $D_{2}$~\cite{Hegger:1999}. The correlation dimension, a phase-space invariant, was estimated via computation of the correlation sum for increasing embedding dimensions of the time series~\cite{Kantz:2004}. The standard embedding procedure of Taken and Sauer was followed using time-delayed reconstruction of the time series~\cite{Takens:1981,Sauer:1991}. The time delay was chosen as the first minimum of autocorrelation to maximize the linear independence of the time delayed vectors. As the phase space was reconstructed from a single time series, a Theiler window of ten times the single round trip time was used to avoid the misinterpretation of temporal correlation as geometrical structure on the attractor~\cite{Theiler:1990}. If the correlation dimension was observed to saturate with increasing embedding dimension the time series was said to have a stable correlation dimension.  If the stable correlation dimension was not an integer then the system was said to be chaotic.

\begin{figure}[ht]
\centering
\includegraphics*[width=.48\textwidth]{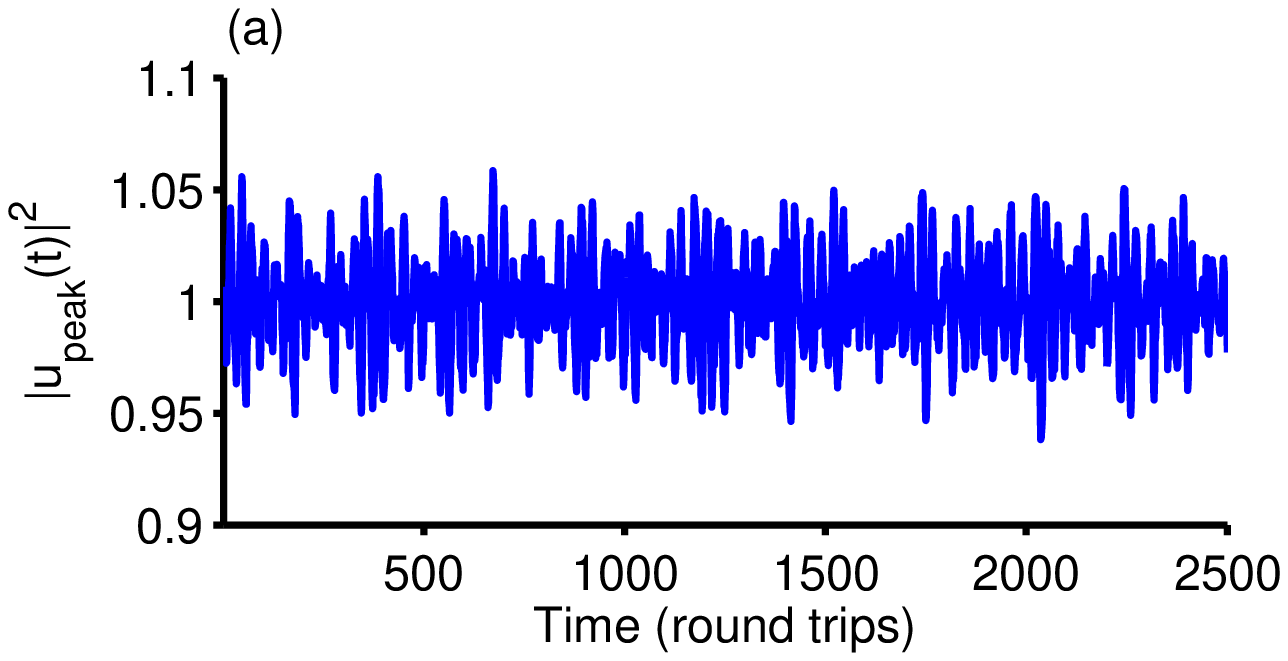}      
\includegraphics*[width=.48\textwidth]{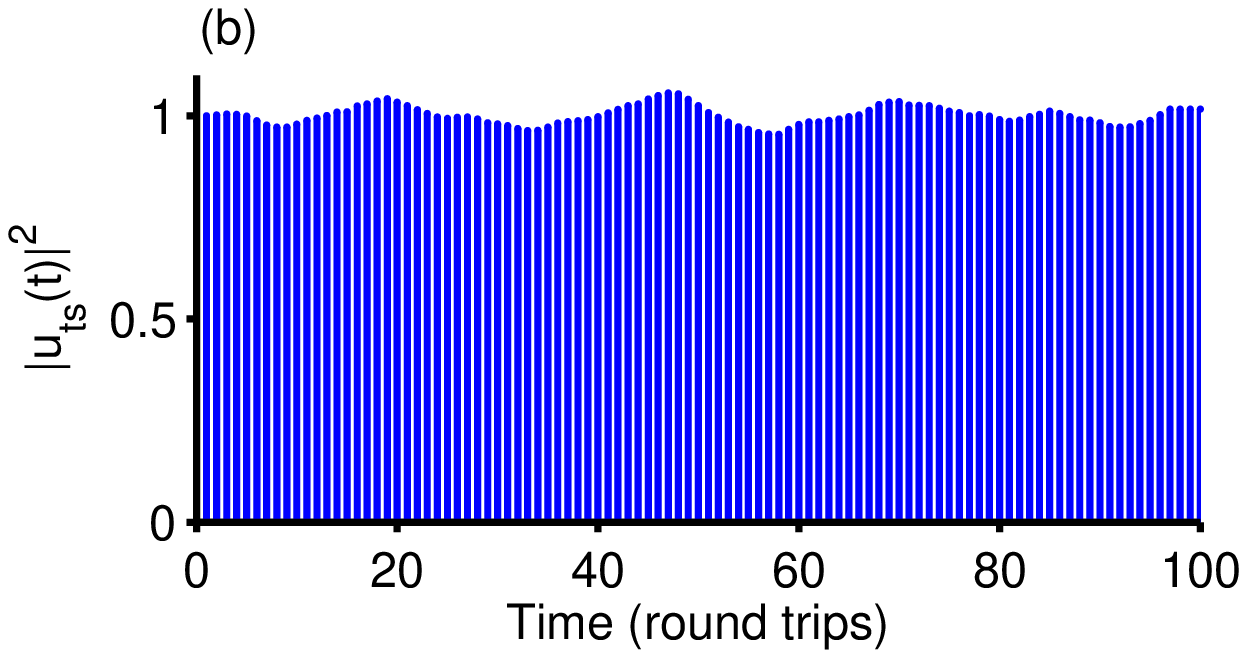}     
\newline
\includegraphics*[width=.48\textwidth]{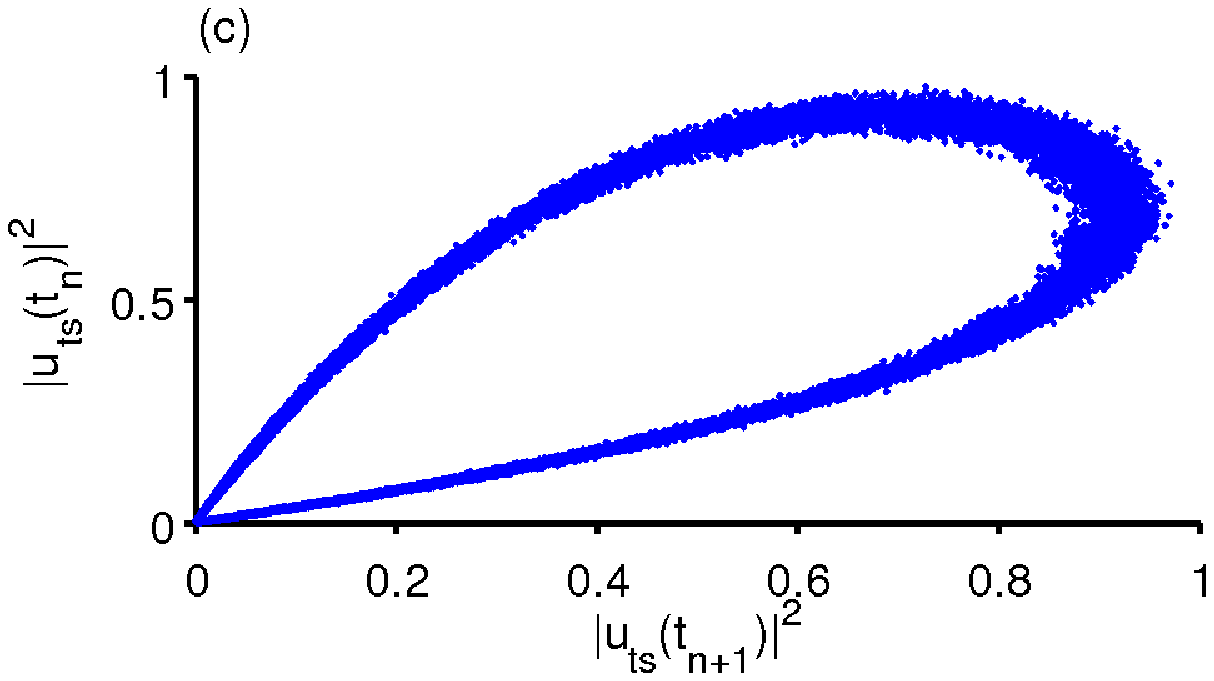}     
\includegraphics*[width=.48\textwidth]{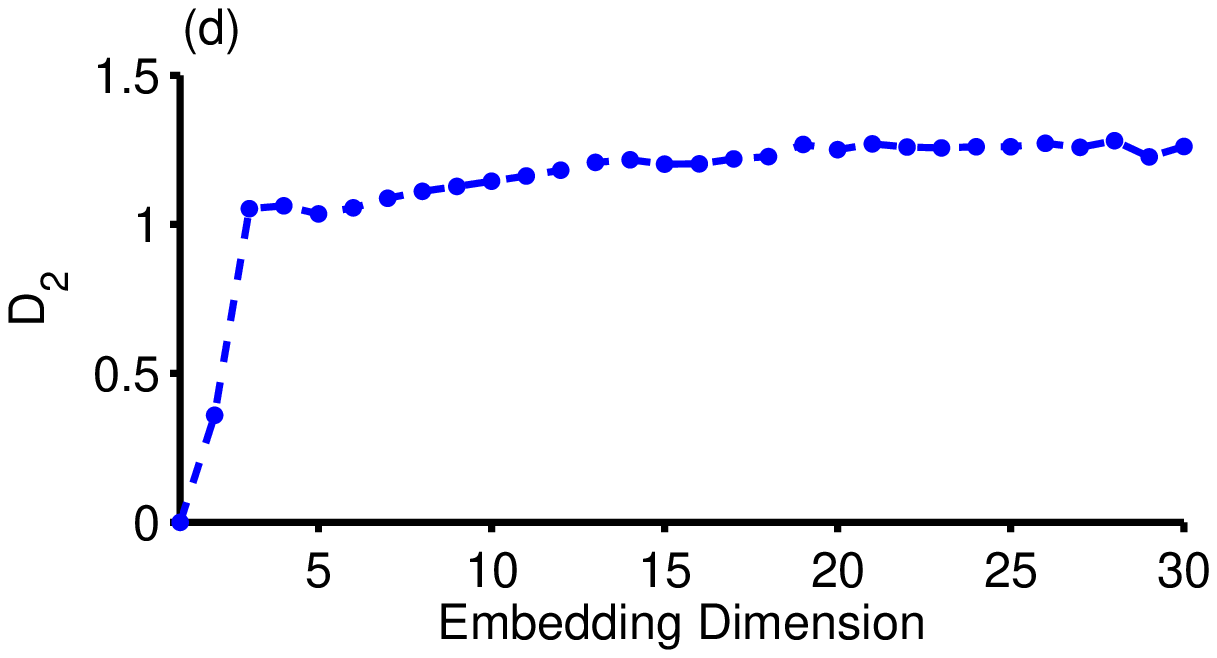}                                        
\caption{\label{fig:chaos1}Low dimensional chaos of a modulated bright soliton train with $2.0\%$ variation about the mean of peak intensity. (a) Peak intensity over 2500 round trips. (b) Time series over 100 round trip; each line shows the bright soliton during a single round trip. (c) Two dimensional return plot of 100000 round trips with a delay time of 1. (d) Correlation dimension vs. embedding dimension with a saturation at $D_2=1.26\pm0.03$. Dashed curve is provided as a guide to the eye; points represent actual data. Reproduced from~\cite{Wang:2011}.}
\end{figure}

We further required the correlation dimension to be stable across a wide range of embedding parameters as one expects the reconstructed attractor to be invariant under smooth transformations.  This requirement was extremely conservative as it was computationally onerous and sensitive to noise. However, such a requirement forbids the optimization of phase space invariants by the tuning of embedding parameters, and the requirement of saturation across embedding dimension eliminates any assumptions required to study a single reconstruction. Additional indicators of chaos include broadband spectra and positive Lyapunov exponents~\cite{Kantz:2004}; note both these properties are shared with noise so a finite correlation dimension is necessary to demonstrate chaotic, rather than random, motion. The principle challenge to finding a stable correlation dimension was isolating a stationary solution.

\begin{figure}[ht]
\centering
\includegraphics*[width=.48\textwidth]{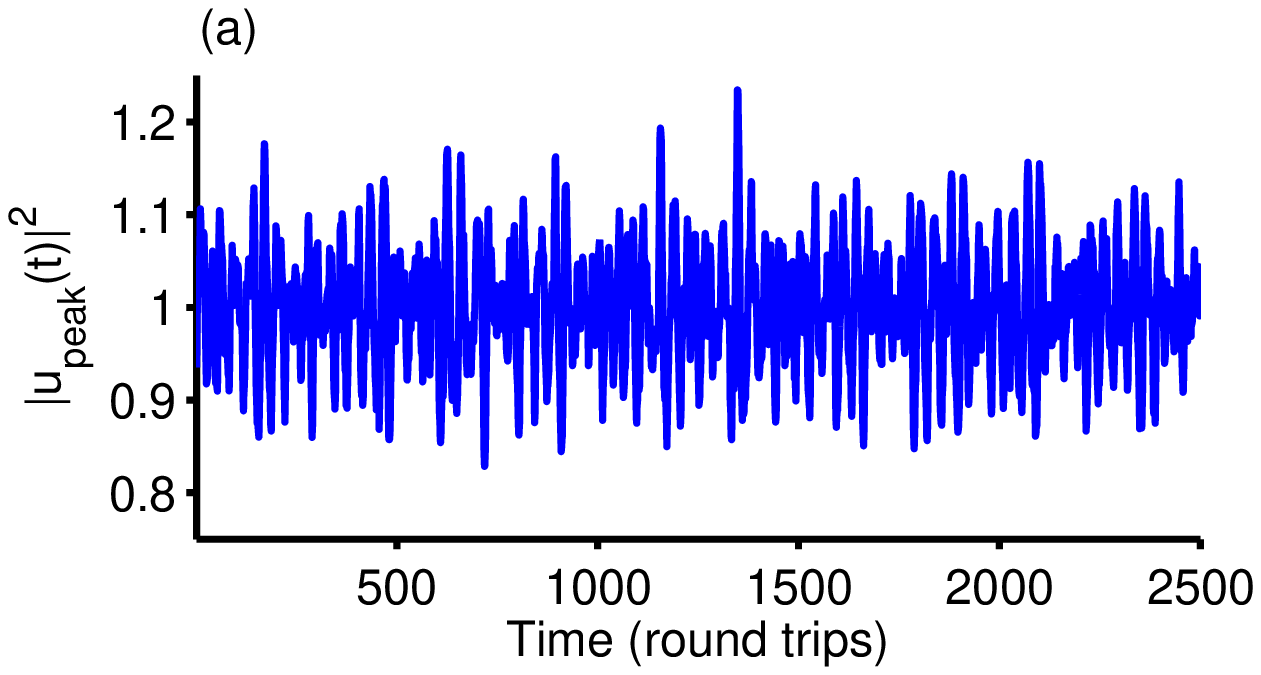}      
\includegraphics*[width=.48\textwidth]{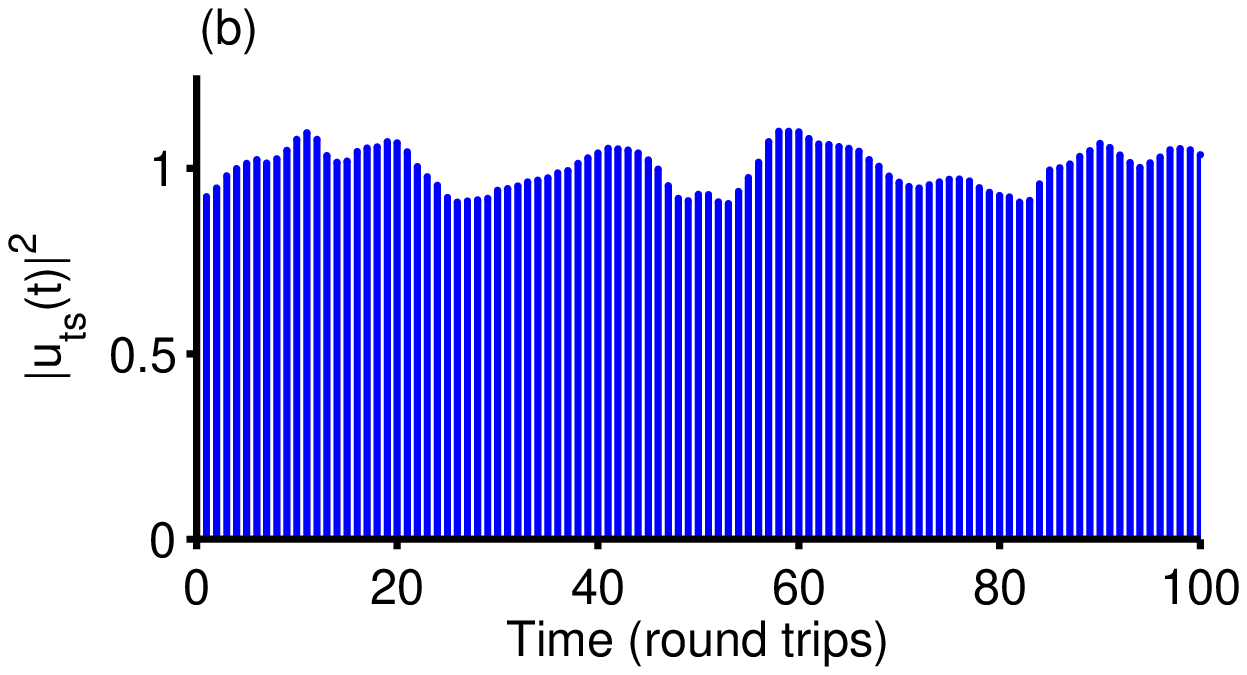}     
\newline
\includegraphics*[width=.48\textwidth]{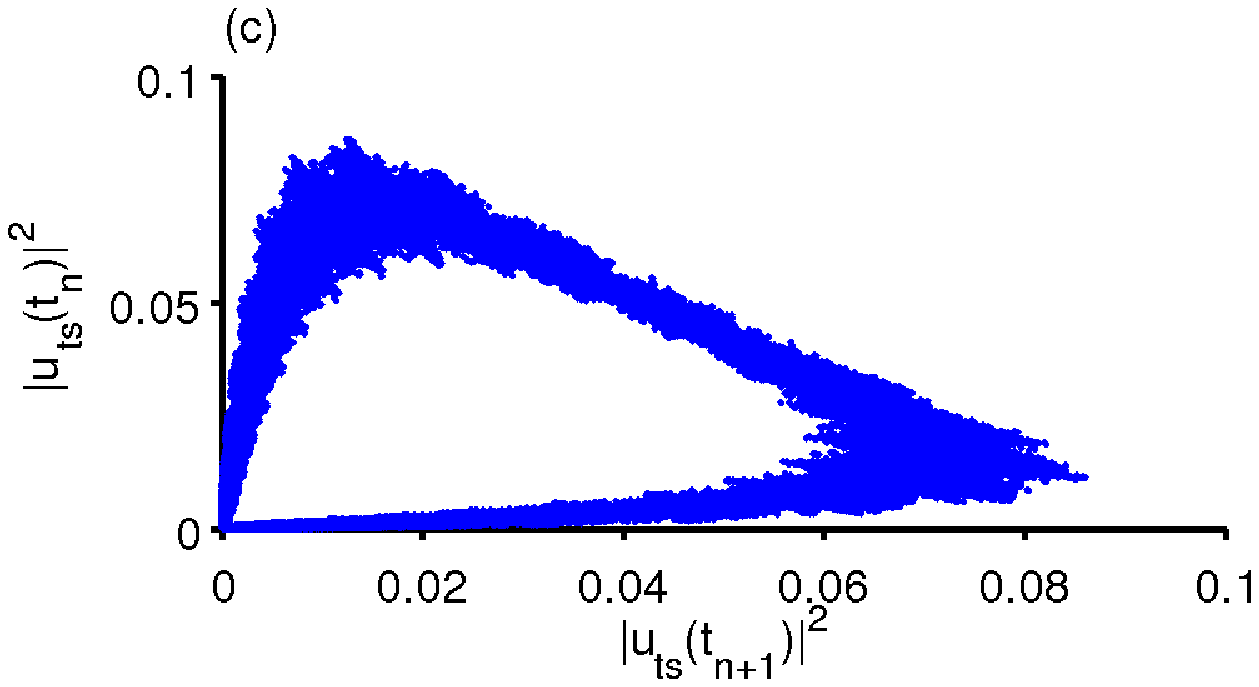}     
\includegraphics*[width=.48\textwidth]{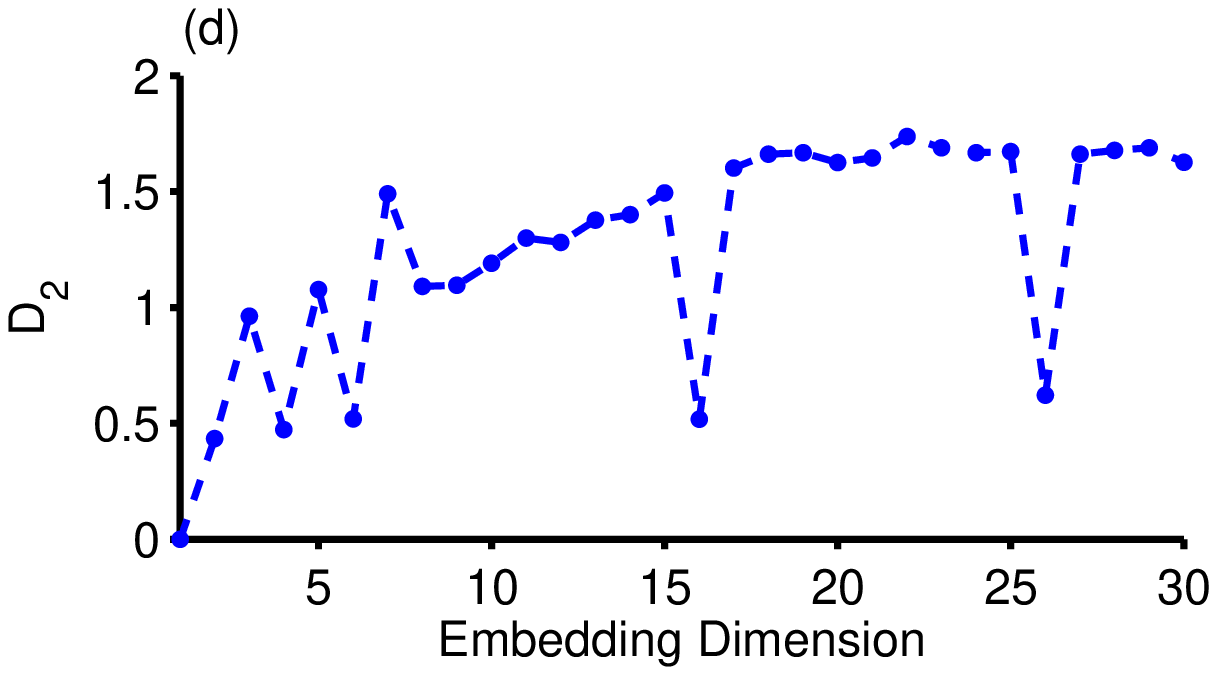}                                        
\caption{\label{fig:chaos2}Low dimensional chaos of a modulated bright soliton train with $5.1\%$ variation of the peak intensity and $D_2=1.66\pm0.07$. Panels treat the same variables as in~\Fref{fig:chaos1}. Dashed curve is provided as a guide to the eye; points represent actual data. Reproduced from~\cite{Wang:2011}.}
\end{figure}

Two examples, with peak variations of $2.0\%$ and $5.1\%$ about their mean, are shown in~\fref{fig:chaos1} and~\fref{fig:chaos2}, respectively. Percent peak variation is defined as
\begin{equation}
100 \frac{\mathrm{var}\left(|u_{\mathrm{peak}}(t)|^2\right)^{1/2}}{ \mathrm{mean}\left(|u_{\mathrm{peak}}(t)|^2\right)},
\end{equation}
where $u_{\mathrm{peak}}$ was previously defined in~\Eref{eqn:upeak} and $\mathrm{var}$ is the sample variance. These values are chosen to match the peak variation of two low ring gain chaotic solitary wave trains observed experimentally by Wang~\etal.~\cite{Wang:2011}. In both figures panel (a) shows the intensity of the single soliton peak for 2500 consecutive round trips while panel (b) shows 100 round trips as would be observed experimentally, as in~\Fref{fig:3a}(d) below, and each vertical line is in fact a bright soliton of finite width. The single soliton peak intensity is immensely complex on inspection and is at worst random and at best chaotic or quasi-periodic.~\Fref{fig:chaos1}(c) and~\ref{fig:chaos2}(c) show the phase space reconstruction for an embedding dimension of 2 and a time delay of 1, also known as a return plot, for 100000 round trips of the full time series. The finite width and structure of the reconstructed attractor is one indication of chaotic, as opposed to random, motion. In~\fref{fig:chaos1}(d) and~\ref{fig:chaos2}(d) is shown correlation dimension versus embedding dimension for each variation case. Both cases saturate above an embedding dimension of 15 to a correlation dimension of $1.26\pm 0.03$ and $1.66\pm 0.07$, respectively. Error estimates are $95\%$ confidence intervals given by two times the standard deviation for values of $D_{\mathrm{2}}$ for embedding dimensions above saturation. This low dimensional chaos closely matches the low ring gain experimental observations by Wang~\etal where $2.0\%$ variation yields a correlation dimension of $D_{\mathrm{2}}=1.27\pm0.12$. However the numerically generated $5.1\%$ peak variation does not reproduce the high dimension chaos, $D_2=3.83\pm0.21$, observed experimentally at matching variations~\cite{Wang:2011}.  The cause of $D_2$ collapse at embedding dimensions 6, 16 and 26 for the $5.1\%$ modulation case has not been rigorously determined but is robust against reasonable perturbations in embedding parameters. The periodicity of the effect suggests the cause is related to sensitivities in the correlation sum to temporal correlations and finite time series. The embedding procedure is also sensitive to time series periodicity, which is present in these low dimensional examples~\cite{Takens:1981,Sauer:1991}. Low dimensional chaos often presents as widened Fourier peaks rather than pure broadband spectra. The oscillation of $D_2$ for low embedding dimension is a common phenomenon as the embedding procedure is not an accurate reconstruction of phase space unless the embedding dimension is at least twice the box counting dimension of the system's attractor~\cite{Kantz:2004}. 

We find numerically that amplitude of peak modulation and the dimensionality of the chaos are principally dependent on the magnitude of the saturating quintic nonlinearity, $Q$. The presence of both a linear gain and nonlinear loss term is necessary for a stable correlation dimension to be determined. Chaotic modulations of the train envelope are the most complex examples of a more general modulation behavior.  Parameter space explorations yielded examples of bright soliton trains with no, periodic, multi-periodic or quasi-periodic modulations.  We note these types of deepening modulations were experimentally observed as the first generations of soliton fractals~\cite{Wu:2006-2}. 

\section{\label{sec:interactions}Symmetric and Asymmetric Interacting Solitary Waves}

When more than one solitary wave propagate with differing group velocities, enabling dynamics such as collisions, we say these waves interact. Two distinct cases where the spatial features of solitary wave interactions are symmetric or asymmetric under rotation are discussed below.

\subsection{\label{sec:interactions:symmetric}Symmetric interaction}

\begin{figure}[ht]
\centering
\includegraphics*[width=1\textwidth]{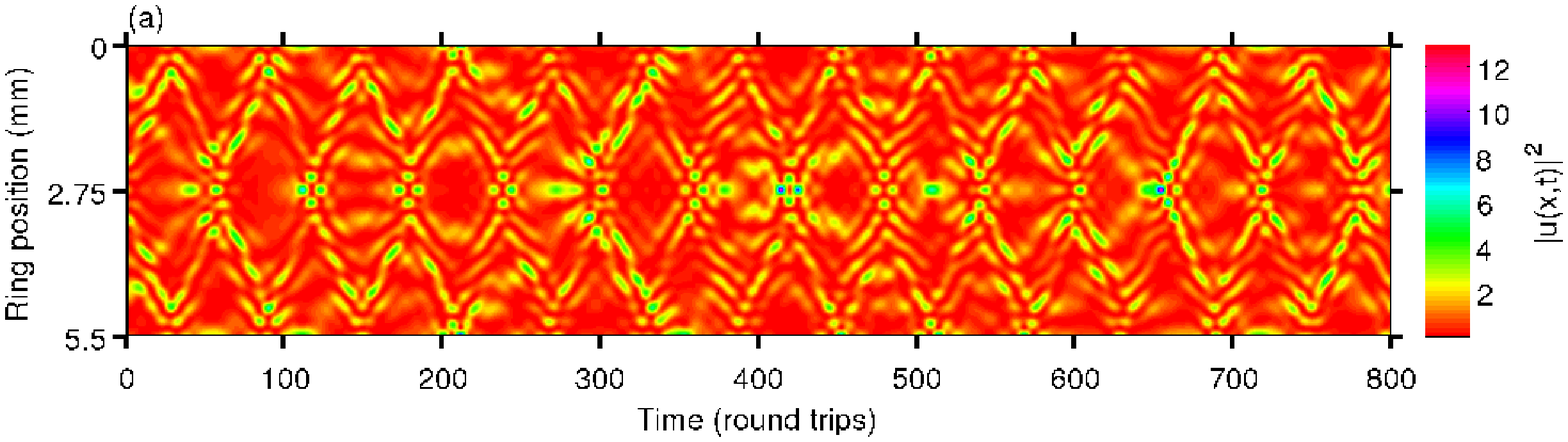}
\newline
\includegraphics*[width=.32\textwidth]{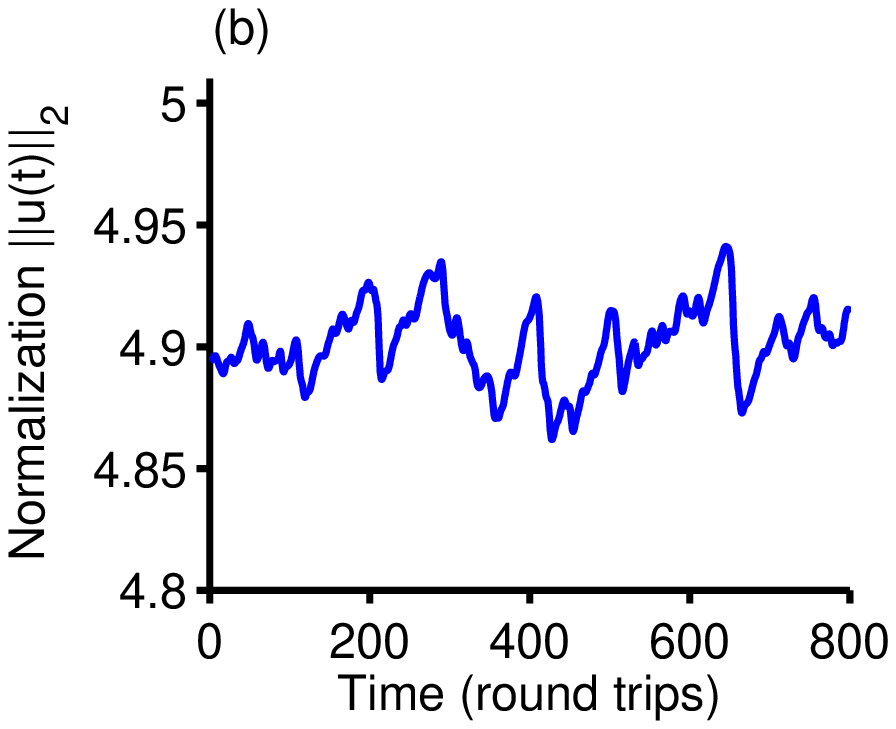}
\includegraphics*[width=.32\textwidth]{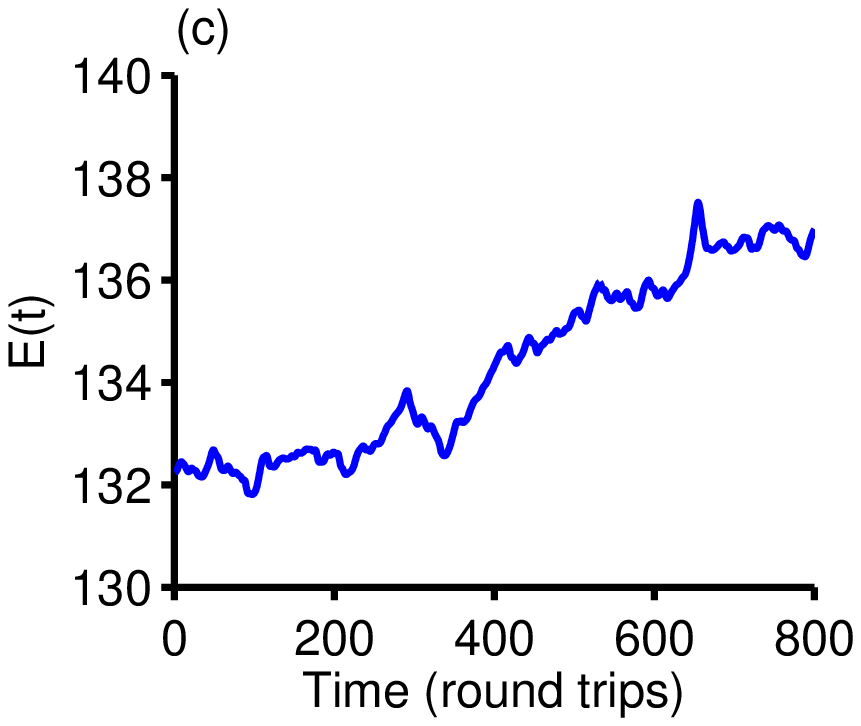}   
\includegraphics*[width=.32\textwidth]{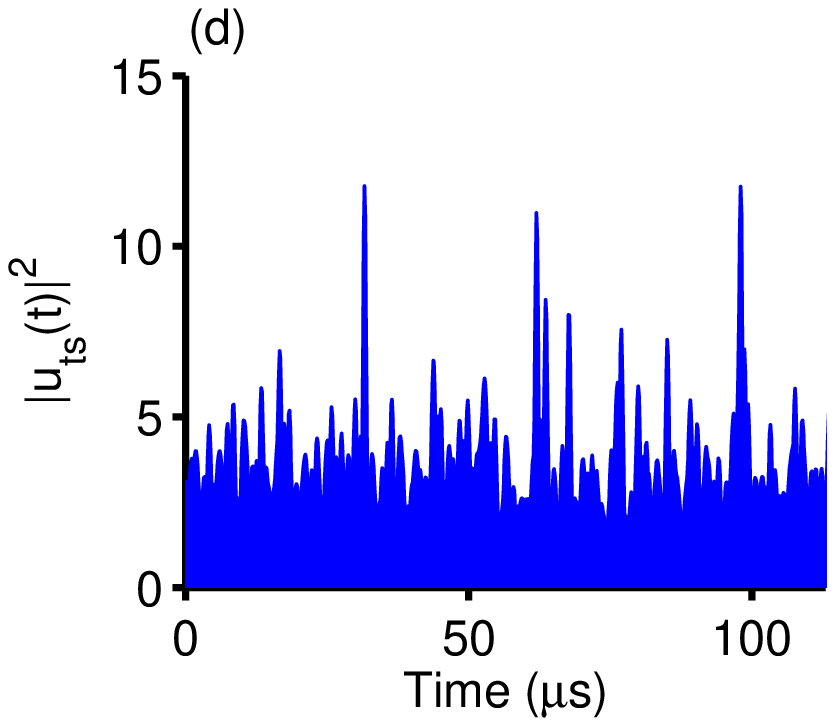}                                                 
\caption{\label{fig:3a}Typical example of a symmetric interaction simulation over 800 round trips of evolution. A stark spatial symmetry about the center of feedback ring, $x=2.75\mathrm{mm}$, is shown in a spatiotemporal plot intensity in panel (a). Symmetric interactions are energetically unstable, see panel (c). Spatial symmetry is not obvious in a reconstructed time series of numerical data, panel (d), which mimics typical experimental data collection (temporally discrete observation at a single point).}
\end{figure}

Symmetric interaction solutions are highly complex, but ordered, gain driven interactions between a number of intensity peaks varying from two to more than twenty. These solutions evolve in intricate and complicated patterns but maintain symmetry in space under a rotation of $\pi$~rads. The solution intensity exhibits a constrained modulation about a stable mean, but is energetically unstable. The energy of the system grows approximately linearly in time and is closely correlated, with a correlation coefficient of $r>.95$, to the time-averaged number of peaks present in the system.  The sample correlation coefficient is a measure of the linear correlation between two variables and is defined as

\begin{equation}\label{eqn:corrcoef}
r(\mathbf{P},\mathbf{E})=\frac{\mathrm{covariance}(\mathbf{P},\mathbf{E})}{\sqrt{\sigma_{\mathbf{P}}\sigma_{\mathbf{E}}}}=\frac{\sum^{n}_{i=1}\left(\mathbf{P_i}-\mathbf{\bar{P}}\right)\left(\mathbf{E_i}-\mathbf{\bar{E}}\right)}{\sqrt{\sum^{n}_{i=1}\left(\mathbf{P_i}-\mathbf{\bar{P}}\right)}\sqrt{\sum^{n}_{i=1}\left(\mathbf{E_i}-\mathbf{\bar{E}}\right)}},
\end{equation}
\noindent where $\sigma_{\mathbf{x}}$ is the standard deviation of $\mathbf{x}$; $\mathbf{P_i}$ and $\mathbf{E_i}$ are the number of peaks and system energy at the $i^{th}$ round trip. This relationship suggests every intensity peak present in the system has similar energy. Peaks undergoing symmetric interactions also demonstrate persistence in time under collisions and have linear or constant phases, both characteristics of bright solitons. Further, individual intensity peaks may also be fit to a $\mathrm{sech}^2$ profile when they are spatially isolated from other peaks circulating the ring. A typical example is illustrated in~\fref{fig:3a}(a) by a spatiotemporal plot of intensity across 800 round trips, each vertical slice shows the waveform on the ring at a specific round trip.  There exists a stark symmetry in dynamics with respect to a rotation by $\pi$ rads.~\Fref{fig:3a}(b) and (c) show the scaled norm and energy, respectively, for the same time frame.  Over these 800 round trips we note the norm varies about a stable mean by $\pm1\%$ while the system energy increases by $8\%$.  A reconstructed time series of data presented in panel (a) is shown in~\fref{fig:3a}(d) to indicate what the behavior would look like if measured experimentally at a single observation point and discretely in time.  We note that the symmetry demonstrated by the spatiotemporal intensity plot,~\fref{fig:3a}(a), is not evident in what appears to be a highly noisy time series.  Whether the symmetry observed numerically persists when the iterative nature of amplification and transmission delays in an electronic feedback loop are considered remains an open question.

Symmetric interactions are observed to evolve in systems with linear gains between $10^{-5}$ and $10^{-3}$ and cubic losses of the same, or $\pm 1$, orders of magnitude.  This long time stable evolution requires a near-balanced system where linear gain is the dominant force and peak growth is meaningfully restricted by the presence of nonlinear losses.  No solutions had initial peak growth above $400\%$ prior to the initial splitting event. 

\begin{figure}[ht]
\includegraphics[width=1\textwidth]{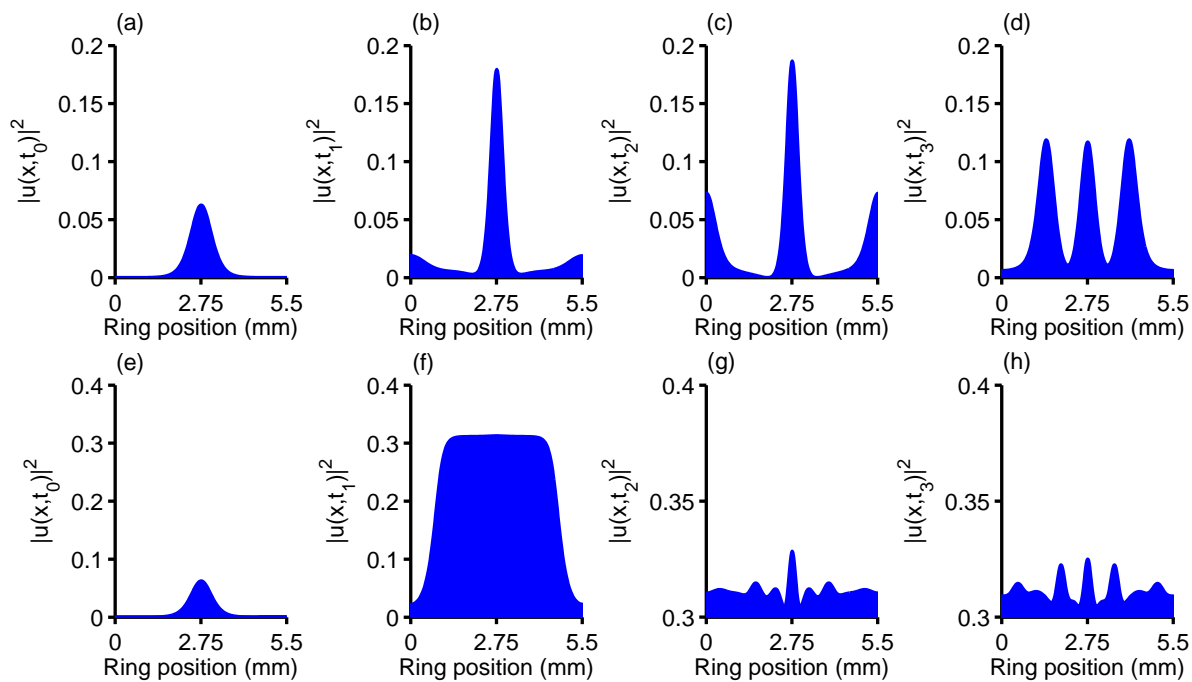}
\caption{\label{fig:splitting_process}A schematic contrast of the splitting process for symmetric interactions, (a)-(d), and asymmetric interactions, (e)-(h).  Each subplot shows the intensity for a single round trip at time $t_0>t_1>t_2>t_3$.  The series progress from left to right. (a)-(d): The symmetric case illustrates a system with linear gain and nonlinear loss near balance, resulting in a slow increase of low intensity regions while peaks are regulated by losses.  Once the floor reaches intensities where nonlinearity affects dynamics, additional solitary wave peaks form, a gain driven process which often takes hundreds of microseconds.  (e)-(h) In contrast the asymmetric system has high linear gains and high nonlinear losses resulting in a flattening of the peak into a plateau with $|m|^2>0$ upon which dynamics occur.  A fast splitting process which occurs on the order of microseconds.  Subplots (g) and (h) have been adjusted to the plateau height. }
\end{figure}

The dynamics within this regime demonstrate a characteristic splitting process, diagrammed in~\fref{fig:splitting_process}(a)-(d).  The initial bright soliton modulates and grows until the domination of linear gain over nonlinear loss in low-intensity regions yields a nonzero intensity floor.  Energy enters the system until these low lying excitations reach intensities where attractive nonlinearity and dispersion may shape the excitation into a stable solitary wave close in form to the well-known hyperbolic secant. The new peak then begins to interact with its neighbors. This same procedure results in the generation of a second, then third, and so on, intensity peak. Thus, in contrast to more typical nonlinear partial differential equations which give rise to fixed soliton dynamics for all times, the GLNLS here displays a particular soliton dynamics on long but not infinitely long time-scales.  This gives rise to the possibility of a new form of integrability which is relevant on long but not infinite times, and may require the development of new mathematical formalisms, in particular a multiscale approach in time. The timing of the initial splitting event varies from $100~\mathrm{\mu s}$ to $1~\mathrm{ms}$ where $t=0$ is defined as the moment gain and losses are turned on. The effect of quintic loss/gain is superficial to this solution category until orders above $10^{-2}$ when it begins to dominate the dynamics.  Quintic losses (gains) result in slower (faster) rates of initial splitting, but do not have any meaningful impact on the rate of energy gain.  This splitting process is stabilized (weakened) by the addition of an attractive (repulsive) quintic nonlinearity term of the same order as the cubic present in the system.  Higher orders of quintic nonlinearity destroy the stability, driving the dynamics into the intermittent regimes described later in~\sref{sec:intermittent}.  This solution type demonstrates a high sensitivity to initial conditions, which is discussed in~\sref{appendix}.  A single round trip of a symmetric interaction solution closely resembles the multi-peaked solitons previously reported by Wu~\etal.~\cite{Wu:2004-2}.

\subsection{\label{sec:interactions:asymmetric}Asymmetric Interaction}

\begin{figure}[ht]
\centering
\includegraphics*[width=1\textwidth]{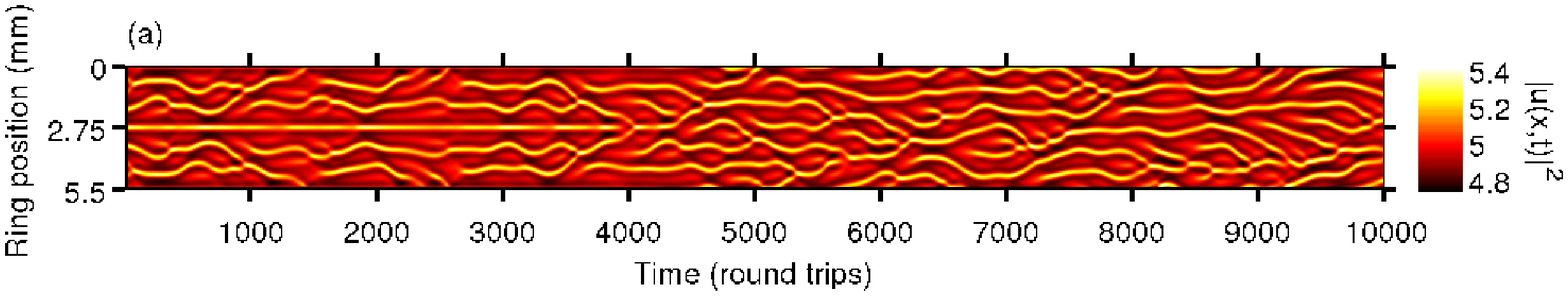}
\newline
\includegraphics*[width=.48\textwidth]{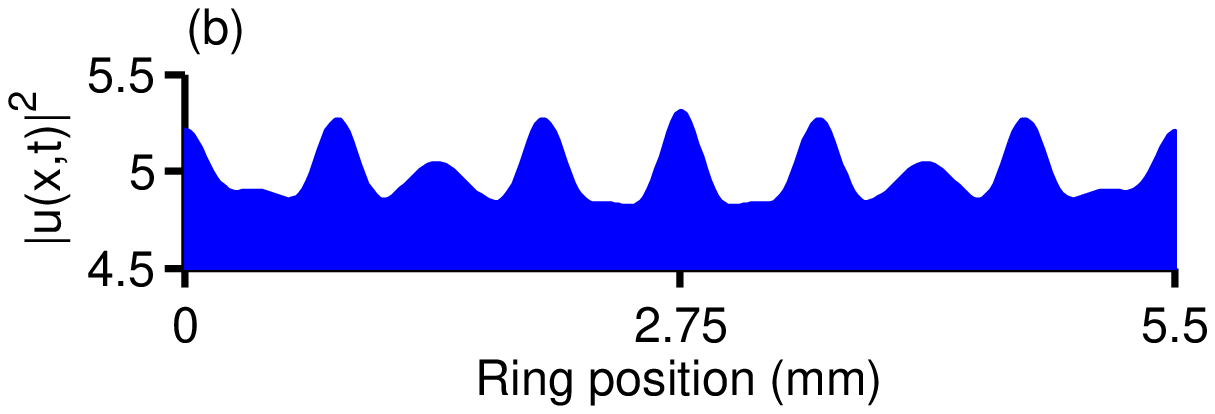}      
\includegraphics*[width=.48\textwidth]{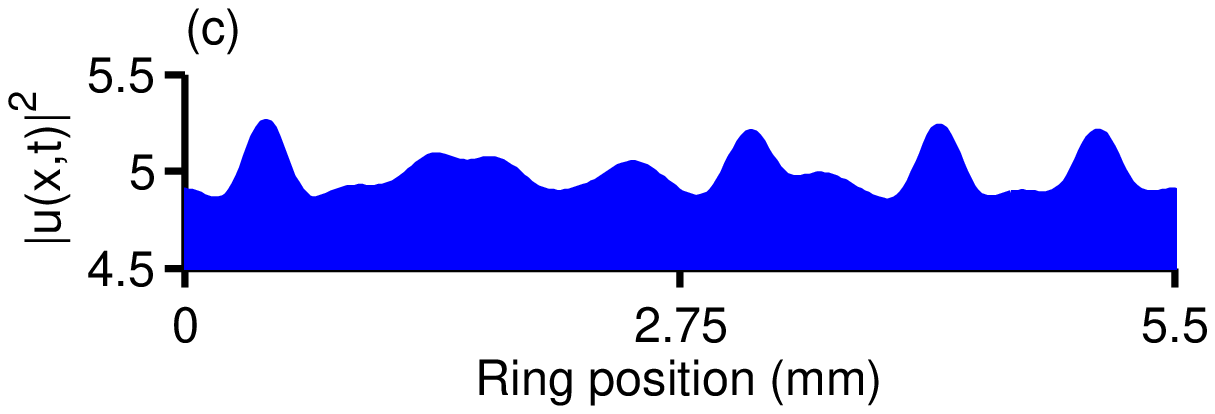}     
\newline
\includegraphics*[width=.48\textwidth]{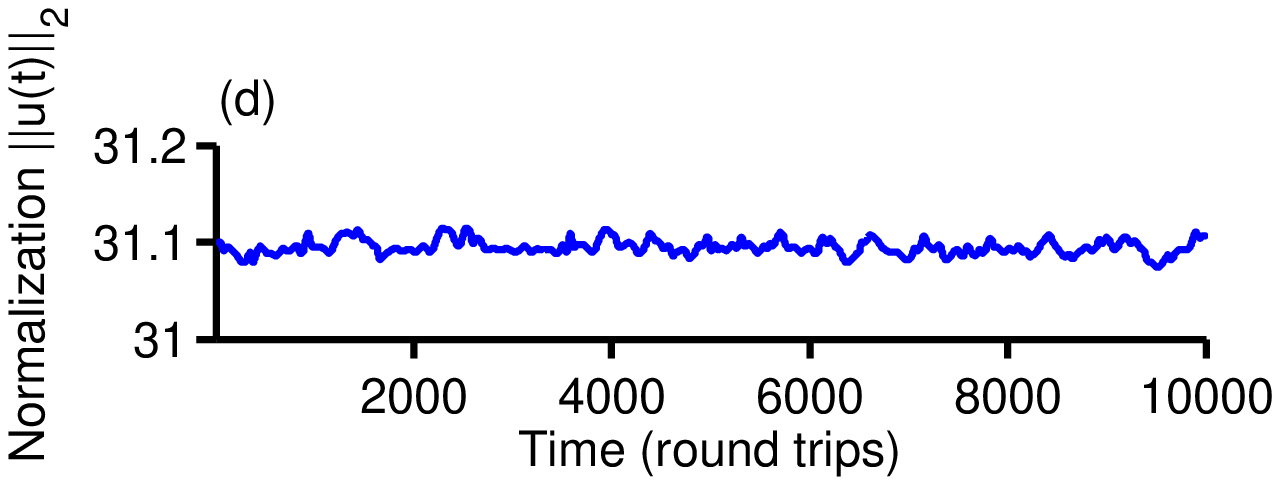}     
\includegraphics*[width=.48\textwidth]{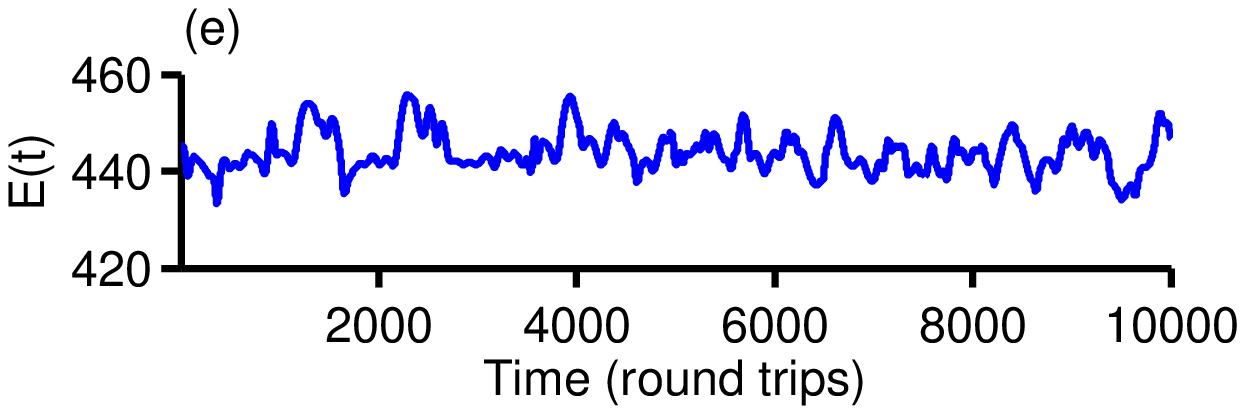}                                          
\caption{\label{fig:3b}Typical example of a asymmetric interaction solution type over 10000 round trips. Spatial symmetry about the feedback ring center, $x=2.75\mathrm{mm}$, can be seen breaking near round trip 4300 in the spatiotemporal intensity plot, panel (a). A spatially symmetric wave form intensity, of round trip 2000, is shown in panel (b) while an asymmetric waveform, round trip 9000, is shown in panel (c).}
\end{figure}

Asymmetric interaction solutions are loss driven solutions which behave similarly to the symmetrical case discussed in~\sref{sec:interactions:symmetric} but do not maintain a spatial symmetry with respect to rotations around the ring.  The number of interacting peaks was observed to vary from five to twenty depending on the parameters of the simulation.  The total number of peaks is conserved, in an average sense, after spatial symmetry about the feedback ring center breaks and is closely correlated, $r>0.98$, to the system's energy.  Here $r$ is the sample correlation coefficient defined by~\eref{eqn:corrcoef}. An example is shown in~\fref{fig:3b}(a) by a spatiotemporal plot of intensity over 10000 round trips.  Symmetry about the feedback ring center, $x=2.75~\mathrm{mm}$, can be seen breaking near round trip 4300. An animation of this symmetry breaking is available online. A scaled intensity plot of a single symmetric (asymmetric) round trip is shown in~\fref{fig:3b}(b) (\fref{fig:3b}(c)).  The interacting peaks are seen to be node-less, and evolve about a non-zero, $|u(x,t)|^2>0$, intensity floor.  The stability of the asymmetric interaction solution type is demonstrated in~\fref{fig:3b}(d) and (e) showing scaled norm and energy, respectively, over the same 10000 round trips.  Normalization varies about a stable mean by $\pm 0.001\%$ while energy modulates by $\pm 3\%$; this stands in contrast to the energetic instability inherent to symmetric interactions,~\sref{sec:interactions:symmetric}.

Asymmetric interactions are observed in systems with linear gains, $L$, between $10^{-4}$ and 1 and cubic loss, $C$, of the same order of magnitude. Quintic gains and losses, $Q$, are stable up to this same order of magnitude. The number of peaks and peak height increased (decreased) with the presence of attractive (repulsive) quintic nonlinearity of the same order as the cubic. Higher orders of quintic nonlinearity push the solution into intermittency, a temporally unstable class of solutions discussed in~\sref{sec:intermittent}.  The solution intensity floor varies with parameter choice, including nonlinearity, but trends towards the constant intensity which satisfies the energy balance of the GLNLS. The balance is given explicitly by the expression

\begin{equation}\label{eqn:energy_balance}
 L+C|u(x,t)|^2+Q|u(x,t)|^4=0\,\,\Rightarrow\,\,|u(x,t)|^2=\frac{-C\pm\sqrt{C^2-4QL}}{2Q},
\end{equation}
\noindent where $L$, $C$, $Q$ and $u(x,t)$ are the same terms as in~\eref{eqn:model:GLNLS} discussed in~\sref{sec:meth} and we choose the smallest positive solution.  For the simulation shown in~\ref{fig:3b} we have $L=0.1$, $C=-0.01$, and $Q=-1$ corresponding to an average solution intensity of $|u(x,t)|^2=0.3113$ which closely matches the numerically observed value of $|u(x,t)|^2=0.3109\pm1.5\times10^{-4}$.  Error bounds are given by two times the standard deviation of intensity across all available round trip data.

This regime demonstrates a characteristic splitting process, diagrammed in~\fref{fig:splitting_process}(e)-(h).  An initial bright soliton initial condition grows and flattens into a plateau under the influence of a strong linear gain and saturating nonlinear losses. Once the non-zero plateau expands to fill the feedback ring the central peak undergoes a splitting procedure similar to that observed for symmetric interactions, diagrammed in~\fref{fig:splitting_process}(a)-(b).  The domination of linear gain over nonlinear losses in low amplitude regions produces small peaks.  These smaller excitations grow until the system's attractive nonlinearity and dispersion shape them into solitary wave intensity peaks.  Unlike the process for symmetric interactions,~\sref{sec:interactions:symmetric}, this splitting process occurs within the first $10~\mathrm{\mu s}$ of evolution, where $\mathrm{t}=0$ is defined as the moment gains and losses are turned on, and saturates within the first $1~\mathrm{ms}$ yielding an energetically stable excitation.  The amplitude of intensity peaks relative to the plateau intensity varies from $1\%$ to $10\%$, but the peak heights measured from the plateau mean are of the same order as those observed in symmetric interactions.

This solution type demonstrates a high sensitivity to initial conditions, which is discussed in~\ref{appendix}. This sensitivity and the highly complex nature of the evolution are hallmarks of chaotic dynamics. However, attempts to arrive at a converged correlation dimension using the methods discussed in~\sref{sec:chaos} were inconclusive. Such sensitivity is typically characterized by a positive Lyapunov exponent~\cite{Kantz:2004}. While a careful determination of the largest Lyapunov exponent requires a rigorous reconstruction of phase space we may estimate the exponent numerically by evolving nearby trajectories in time. Direct measurement suggests a Lyapunov exponent between $\lambda=2\times 10^4~s^{-1}$ and $\lambda=1\times 10^5~s^{-1}$. This rate of trajectory separation is of the same order as that observed experimentally ($\lambda = 1.9\pm0.2\times 10^5 s^{-1}$~\cite{Wang:2011}) for the $5.1\%$ modulating soliton train discussed previously in~\sref{sec:chaos}.

\section{\label{sec:DPF}Dynamical Pattern Formation}

Four distinct robust dynamical patterns which demonstrate lifetimes of at least $1~\textrm{ms}$ or 7000 round trips were located during GLNLS parameter space exploration.  Solutions of this group differ from previously discussed solution behaviors in that they exhibit a periodic recurrence of their characteristic dynamic. Self organization of this kind is common in open nonlinear systems~\cite{Cross:1993}.  These examples are discussed to demonstrate the breadth of pattern formation supported by the GLNLS under fixed choice of $N$ and $D$.  The regions of parameter space supporting dynamical pattern formation violates the assumptions underlying the derivation of the GLNLS in the context of magnetic spin waves, as discussed in~\sref{sec:meth}, owing to the high order of quintic nonlinearity and losses which drive evolution. However, the GLNLS is a useful model in a variety of systems including laser cavities, as discussed in~\sref{sec:intro}, and these dynamics may appear in such contexts.

\subsection{\label{sec:DPF:recombination}Central Peak Recombination}

\begin{figure}[ht]
\centering
\includegraphics*[width=.65\textwidth]{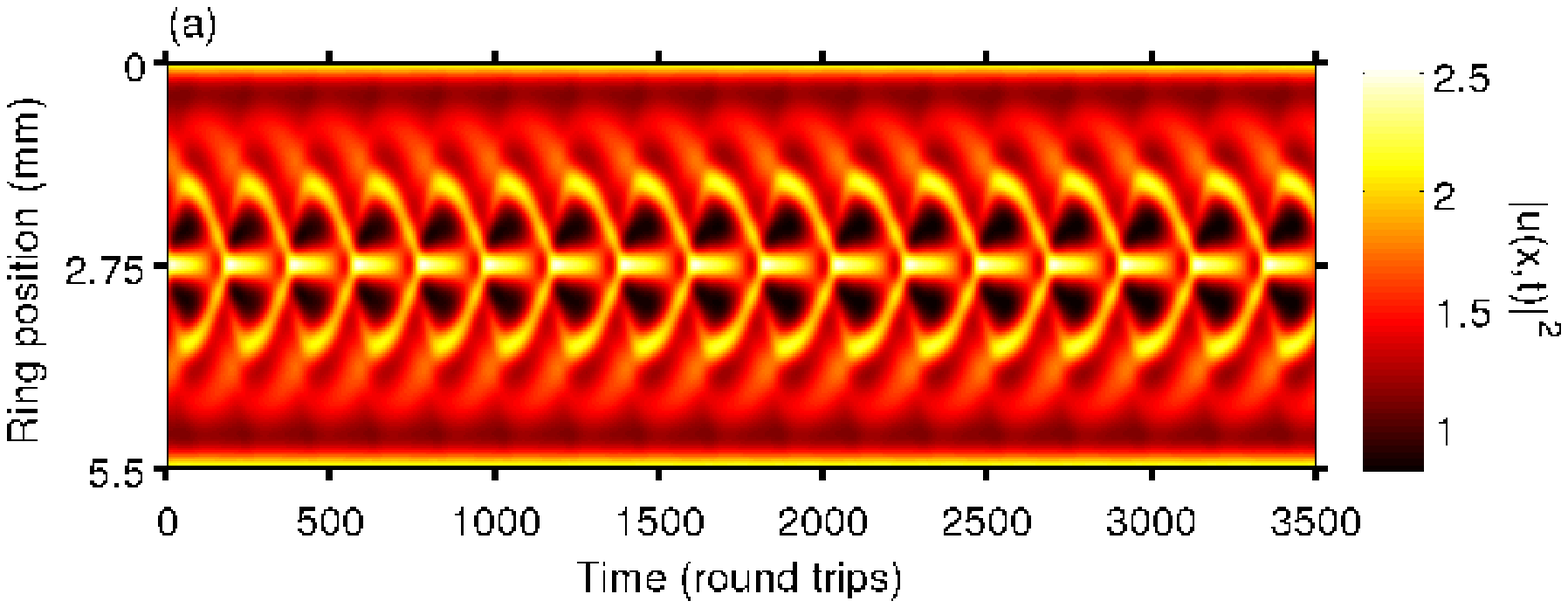}
\includegraphics*[width=.32\textwidth]{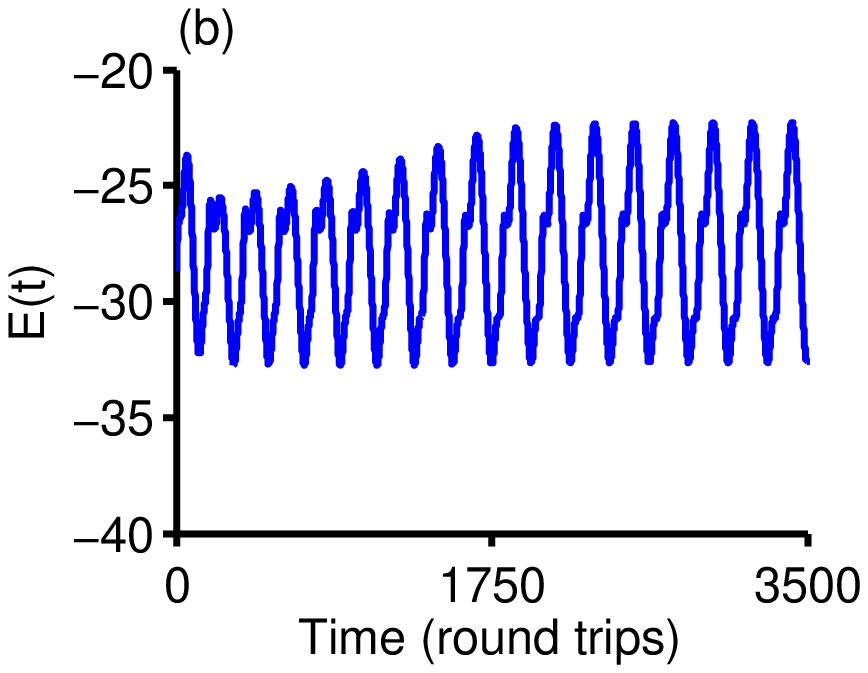}
\\
\includegraphics*[width=.65\textwidth]{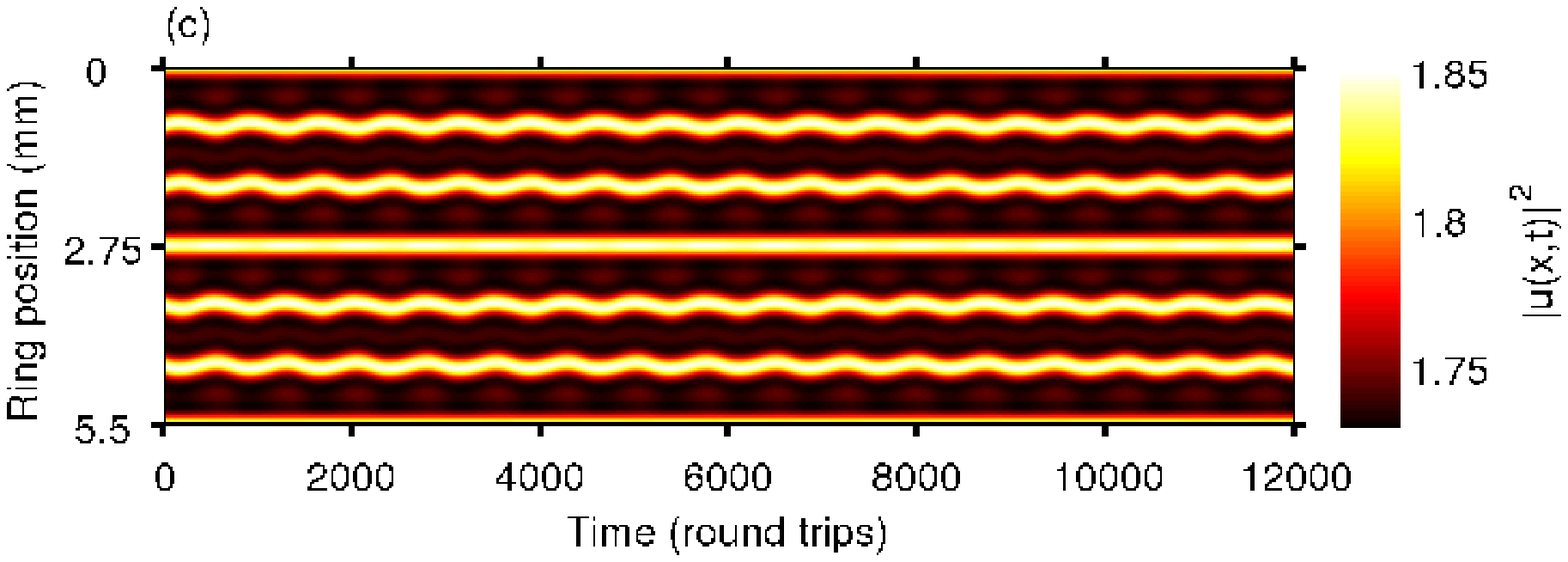}
\includegraphics*[width=.32\textwidth]{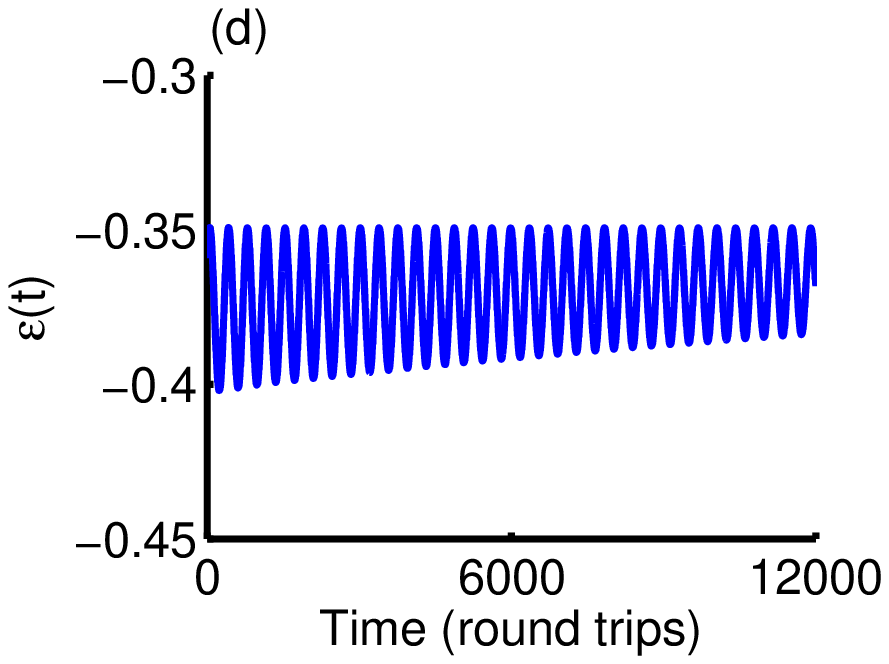}
\caption{\label{fig:CCP_DPF}Examples of dynamical pattern formation with experimentally observable lifetimes. (a)-(b) Central peak recombination. (c)-(d) Complexly co-propagating solitary waves. In panel (d) energy has been offset for clarity, $\epsilon(t)\equiv\left(E(t)+142\right)$.}
\end{figure}

Central peak recombinations exhibit a complex 5 peak solitary wave recombination pattern with a periodicity of 180-250 round trips, depending on parameter choices. This behavior is driven by a strongly attractive quintic loss, $Q=-1$ and a linear gain of $L=10^{-2\pm 1}$ with quintic loss of $Q=-10^{-3\pm 1}$.  The presence of quintic nonlinearity has a severely negative impact on the behavior lifetime. The median wave height of central peak recombination solutions satisfies the energy balance equation,~\eref{eqn:energy_balance}. For the example shown in~\fref{fig:CCP_DPF} we predict an average intensity of $|u(x,t)|^2=0.0995$, corresponding to the parameters have $L=0.01$, $C=-0.001$ and $Q=-1$, which closely matches the observed numerical average intensity, $|u(x,t)|^2=0.0934\pm6\times10^{-3}$. The error estimate is defined as in~\sref{sec:interactions:asymmetric}.

An example of central peak recombination is shown in~\fref{fig:CCP_DPF}. Panel (a) shows a spatiotemporal plot of scaled intensity over 3500 round trips and (b) shows the scaled energy over the same round trips. The periodicity of the recombination is evident in the spatiotemporal plot, which contains 16 periods. The dynamics are most readily described starting when the central peak collapses. Immediately following the collapse, the peaks on either side of the ring center propagate towards the middle of the ring and recombine into a new central peak matching the original peak's amplitude. At the same time the outlying peaks split into two. The innermost of these new peaks grows until one finds three central peaks of equal amplitude. At this point the central peak undergoes collapse and the process repeats.  A single bright solitary wave propagates unperturbed along the edge of the ring. This process is animated in an attached movie, available online.

\subsection{\label{sec:DPF:complexcoprop}Complex Co-propagation}

The complex co-propagation solution was so named as it resembles the steady state co-propagation solution (see ~\sref{sec:SS:coprop} below) and is likewise energetically stable. It differs primarily in that the waveform undergoes complex, but periodic, modulation. The dynamics also occur on a non-zero density floor satisfying the GLNLS energy balance equation,~\eref{eqn:energy_balance}. The example shown in~\fref{fig:CCP_DPF}(c)-(d) was simulated with the parameters $L=0.0987$, $C=-0.0505$, $Q=-7.6261$, resulting in an anticipated average intensity of $|u(x,t)|^2=0.1105$. This prediction closely matches the numerically observed intensity $|u(x,t)|^2=0.1104\pm4\times10^{-7}$, where the error is defined as in~\sref{sec:interactions:asymmetric}. Like central peak recombination the complex co-propagation behavior is driven by a large quintic loss. The dynamical patterns demonstrated by these solutions also require a large attractive quintic nonlinearity.  The parameter space region which supports these behaviors is characterized primarily by large, negative quintic terms: $S=Q\ge-1$. The smallest linear gain which compensates these driving nonlinear losses is $L=0.01$.  These solutions are in general insensitive to the choice of cubic loss, with any value smaller than $C=-0.1$ supporting the observed dynamical pattern.

\Fref{fig:CCP_DPF}(c)-(d) illustrates this behavior.  Panel (c) shows a spatiotemporal plot of scaled intensity over 12000 round trips and (d) shows the scaled energy over the same round trips. In panel (d) the energy has been offset for clarity, $\epsilon(t)\equiv\left(E(t)+142\right)$. The behavior is characterized by the spatiotemporal plot which shows two spatially stable bright solitary waves occupying the center and edges of the ring.  The central solitary wave is flanked on each side by a set of two periodically oscillating solitary waves for a total of six large peaks being equispaced around the ring. Six additional small amplitude peaks occupy the space between each larger wave. The entire waveform breathes between two distinct energy states with a period of 750 round trip times. The frequency of oscillation matches that predicted by a simple two-level quantum system where $\omega=\Delta E / \hbar$. For the GLNLS we have $\hbar=1$ amd $t=25~\mathrm{ns}$, as defined in~\sref{sec:meth}. Taking the average energy difference between states, see~\fref{fig:CCP_DPF}, one predicts an angular frequency of $\omega=7900 s^{-1}$ compared to the observed oscillation frequency of $\omega=8100 s^{-1}$. An animation of the breathing is available online.

\begin{figure}[ht]
\centering

\includegraphics*[width=.65\textwidth]{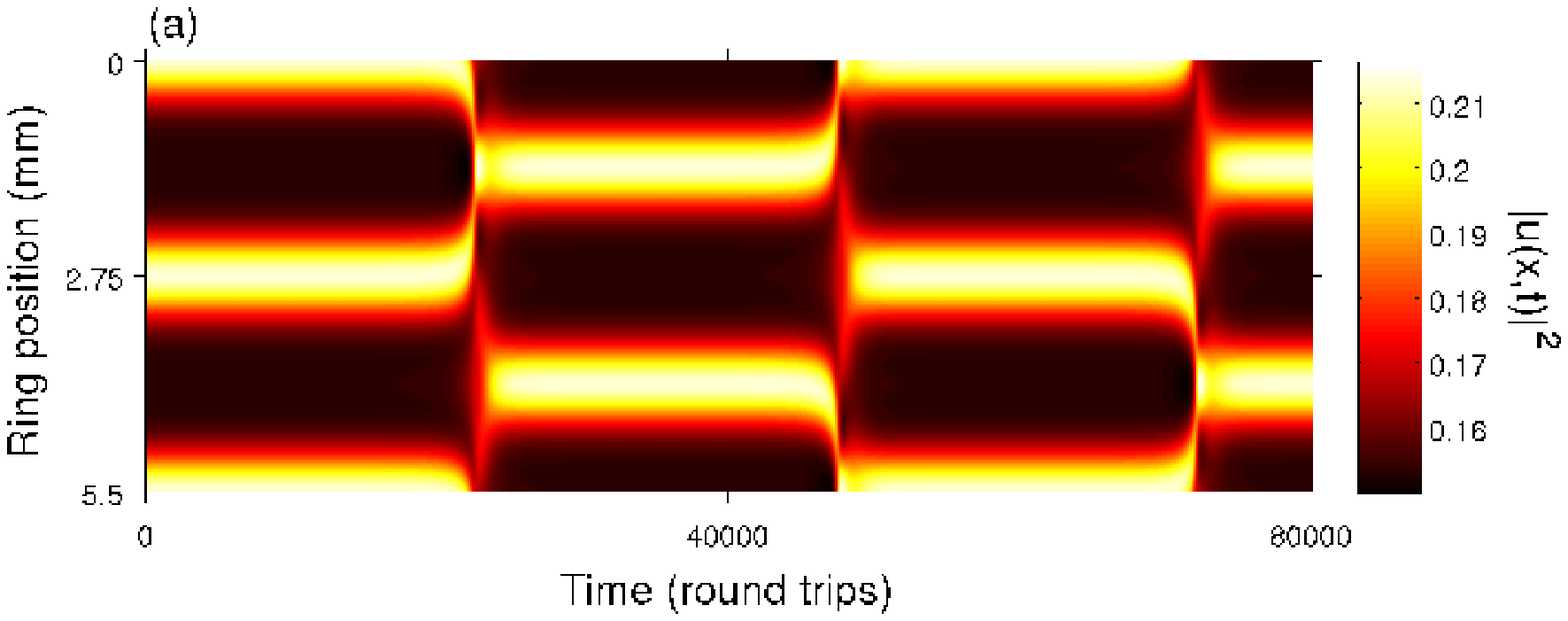}
\includegraphics*[width=.32\textwidth]{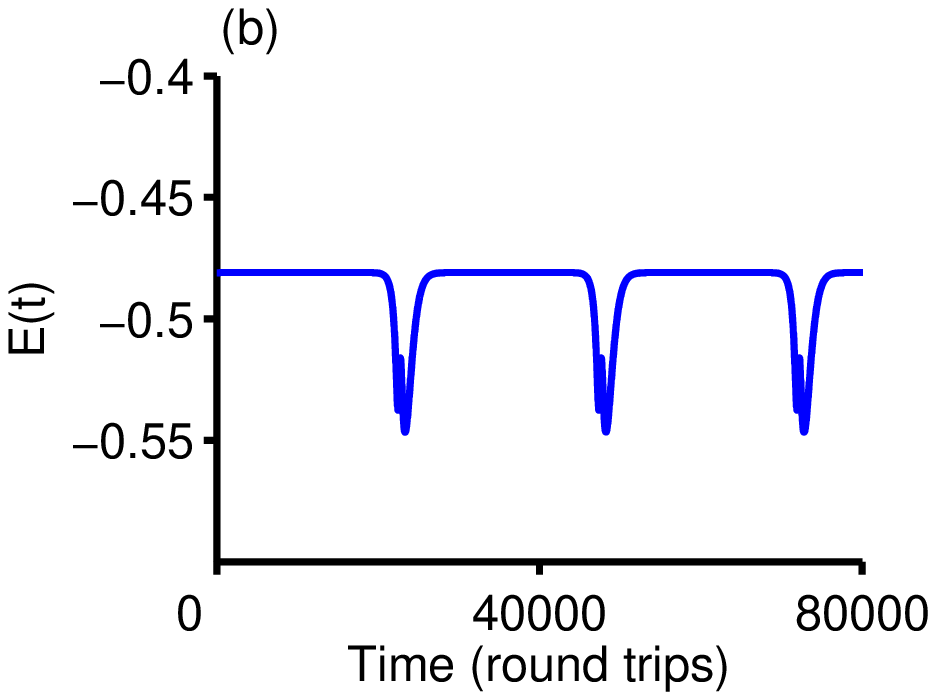}
\\
\includegraphics*[width=.65\textwidth]{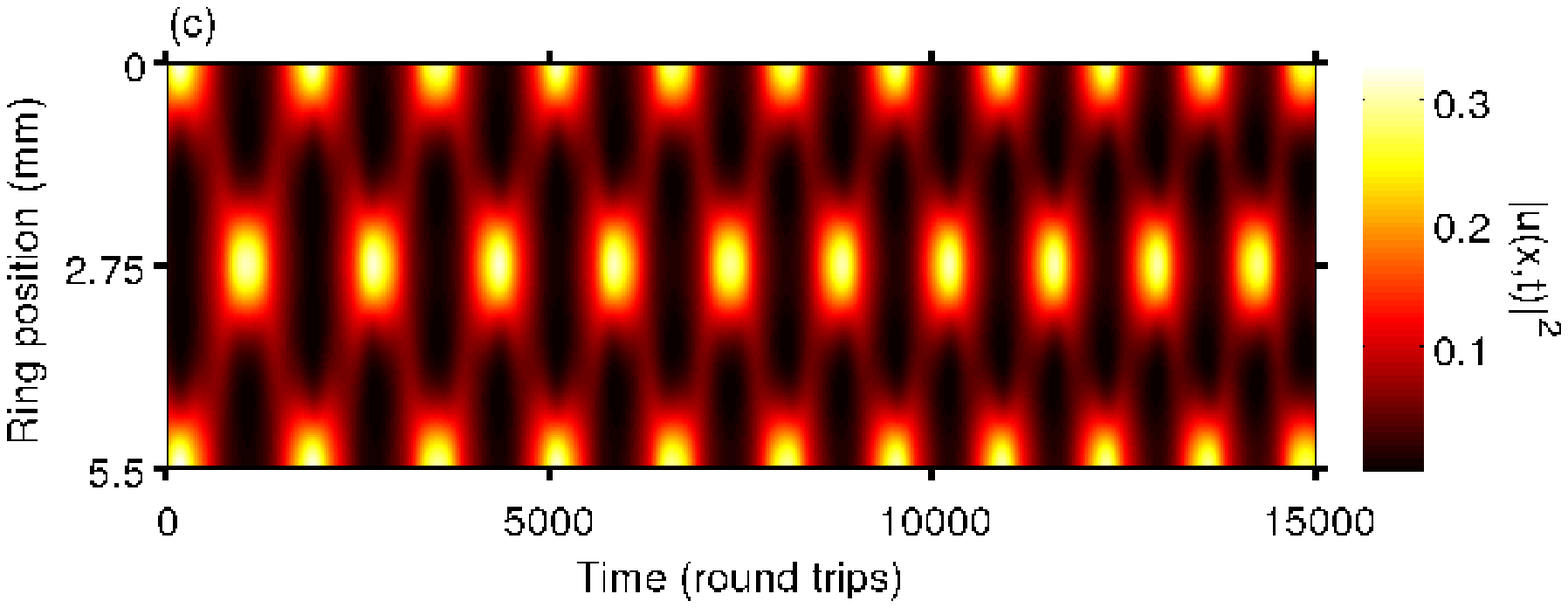}
\includegraphics*[width=.32\textwidth]{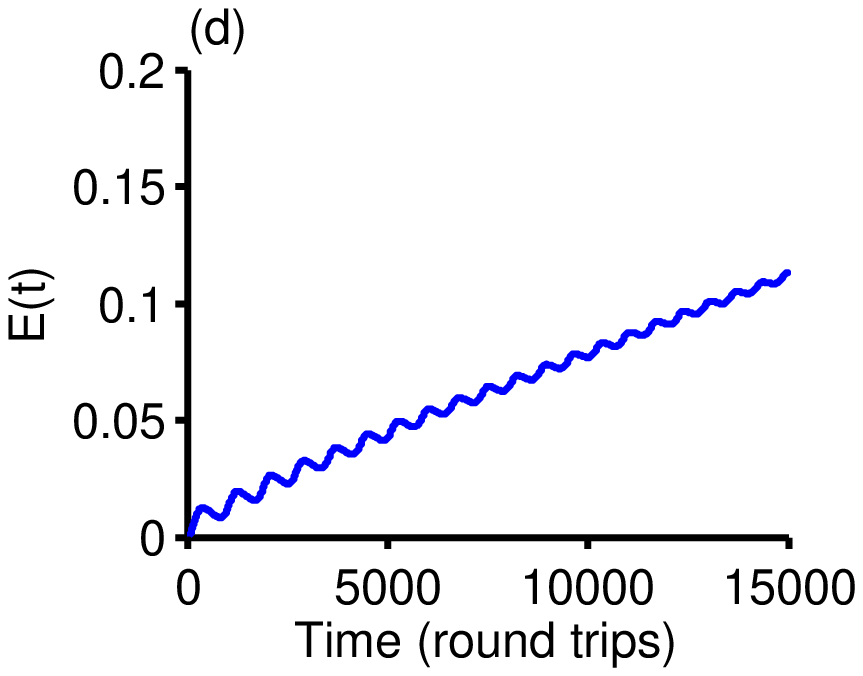}
\caption{\label{fig:EC}Examples of dynamical pattern formation. (a)-(b) Spatially shifting solutions. (c)-(d) Egg carton solutions.}
\end{figure}

\subsection{\label{sec:DPF:spaceshift}Spatial Shifting}

Spatial shifting solutions are simulations which exhibit energetically stable evolution with a well-defined and periodic shifting of the spatial location of the dynamical behaviors. In all observed cases the underlying energetically stable dynamics are evenly distributed bright solitary waves co-propagating on an intensity floor which satisfies the GLNLS energy balance given by equation~\eref{eqn:energy_balance}. For the example shown in~\fref{fig:EC}(a)-(b) we have $L=0.02899$, $C=-0.06219$ and $Q=0.000648$ corresponding to $|u(x,t)|^2=0.0438$ which closely matches the numerically observed average intensity of $|u(x,t)|^2=0.0432\pm2\times10^{-4}$. The solitary waves spontaneously split at a constant periodicity and reform into an identical set of co-propagating peaks with a spatial shift defined as $\frac{L}{2N_s}$ where $L$ is the feedback ring length and $N_s$ is the number of peaks present in the simulation.  All peak properties as well as the splitting dynamics remain consistent through multiple periods. A strong attractive quintic nonlinearity is required to support this dynamical behavior, as seen previously with central peak recombinations and complex co-propagation in sections~\ref{sec:DPF:complexcoprop} and~\ref{sec:DPF:recombination}.  Spatial shifting is seen in simulations with quintic nonlinearities of $Q\simeq-0.8$ and moderate linear gains of $L=10^{-2\pm 1}$. Cubic losses near $C=10^{-1}$ support this behavior, while quintic losses were found to be unimportant until above values of $Q=\pm 10^{-1}$ where they dominated the dynamics.  

An example of temporal shifting is illustrated in~\fref{fig:EC}(a)-(b).  Panel (a) shows two bright solitary waves co-propagating while undergoing a spatial shift of $\frac{5.5}{4}~\mathrm{mm}$ every 22000 round trips. The shifting event occurs over 1500 round trips. Panel (b) shows the solution's scaled energy over these same round trips; the energetic stability of the co-propagation regimes is demonstrated.  The energy profile of each shifting event was found to be identical.

\subsection{\label{sec:DPF:breather}Breathers}

Solitary wave breathers on a ring are characterized by a single solitary wave which undergoes a periodic disappearance of the peak and reappearance at the other side of the ring. The frequency of breathing increases with system energy. The solution is not energetically stable and breathing frequency increases until the system reaches a new dynamical behavior. Numerically observed lifetimes were never less than 20000 round trips, or 3 milliseconds. The wave breathing is driven by a strong quintic loss, $Q=10^{-1}$, with comparatively weak linear gain, $L=10^{-5\pm 1}$, and cubic loss, $C=10^{-4\pm 1}$, terms.  The solution type is sensitive to the presence of quintic nonlinearity with any magnitude above $10^{-2}$, whether attractive or repulsive, pushing the dynamics into the intermittent regime, discussed later in~\sref{sec:intermittent}. Linear gain dominates during low intensity periods of the breathing behavior, resulting in a non-zero intensity floor which ultimately drives the collapse of stable breathing. 

\Fref{fig:EC}(c)-(d) contains a typical example of solitary wave breathing. The periodic spatial shifting of the bright solitary wave is seen as a relocation from the center of the ring to the other side in panel (c). An average breathing period of 1200 round trips is observed in this example.  A positive linear trend in energy, see panel (d), is the result of linear gain causing growth in low-intensity regions. A periodic high rate of energy growth matches the low intensity period following the collapse of bright solitary waves.  An animation of the breathing behavior is available online.

\section{\label{sec:steady_state}Steady State Solutions}

Simulations which evolved into energetically stable static wave forms were named steady state solutions.  Two distinct steady state solutions were isolated from the parameter space exploration: multi-peaked solitary waves and co-propagating solitons.
  
\begin{figure}[ht]
\centering
\includegraphics*[width=1\textwidth]{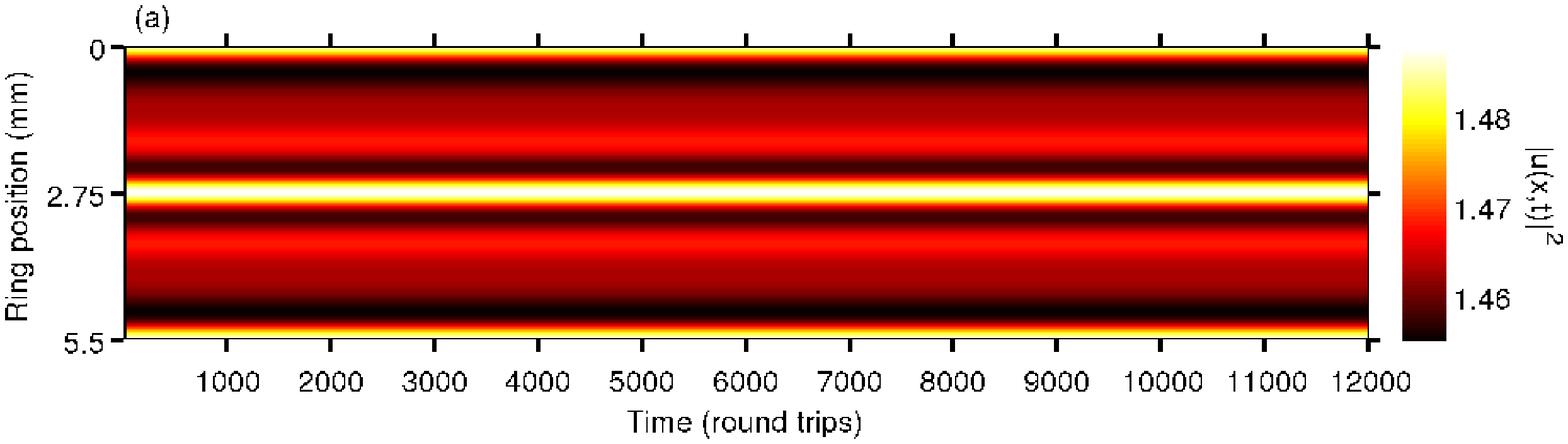}
\\
\includegraphics*[width=.48\textwidth]{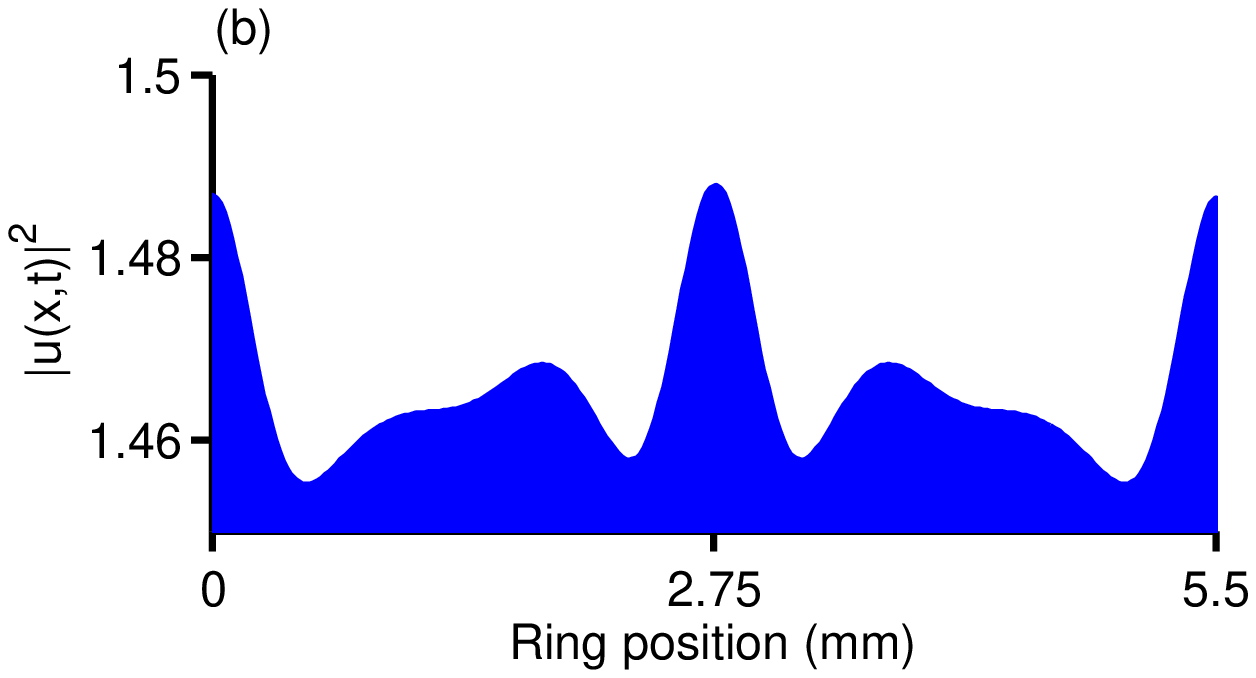}     
\includegraphics*[width=.48\textwidth]{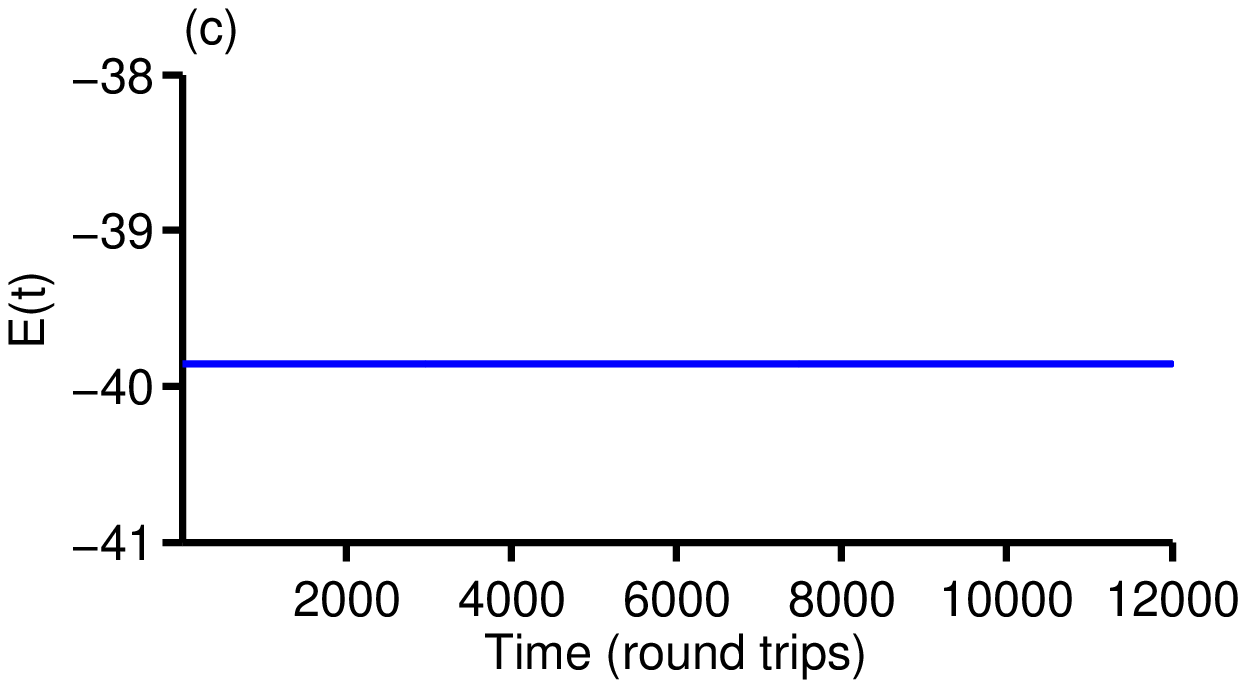}  
\caption{\label{fig:3d}Typical example of a multi-peaked soliton solution type with $N_s=2$ principle solitary wave peaks.}
\end{figure}

\subsection{\label{sec:SS:multi}Multi-peaked Solitary Waves}

Multi-peaked solitary waves were characterized by energetically stable, to machine precision, nodeless complex waveforms that evolve without exhibiting any time dependence in their intensity.  The shape of the wave and the number of principle peaks varies from two to eight in studied cases, depending on parameter choice. Symmetric and asymmetric waveforms were observed. Multi-peaked solitary waves were observed for any linear gain, $L$, below 1 and cubic losses, $C$, of $\pm 3$ orders of magnitude.  The impact of quintic losses and gains principally affected the median wave height according to the GLNLS energy balance equation,~\eref{eqn:energy_balance}. For the multi-peaked solitary wave shown in~\fref{fig:3d} we have $L=0.1$, $C=-1$ and $Q=-1$ corresponding to an estimated average intensity of $|u(x,t)|^2=0.0916$ which closely matches the observed value, $|u(x,t)|^2=0.0916\pm7\times10^{-7}$. The error was previously defined in~\sref{sec:interactions:asymmetric}. As with previous examples, high values of $Q$ relative to $L$ lead to the term dominating dynamics and the solution leaving the steady state solution class. Positive, or saturating, values of quintic nonlinearity lead to reductions in secondary peak heights while attractive values leads to the presence of additional principal peaks via further shaping of secondary peaks. The overall shape of the multi-peaked solitary wave, including the number of principal and secondary peaks, is dependent on the choice of parameters.  

A typical example is shown in~\fref{fig:3d} of a symmetric multi-peaked solitary wave with two principle and two secondary peaks.~\Fref{fig:3d}(a) shows a spatiotemporal plot of scaled intensity over 12000 round trips with each vertical slice showing the intensity across a single round trip. Panel (b) is the scaled intensity plot of the final round trip and panel (c) shows the static solution energy over the same evolution period.  Not all multi-peaked solitary waves travel at the group velocity as the example in~\fref{fig:3d}. This solution type is the most commonly observed long time behavior in studied simulations and was one of the behaviors present in a majority of the intermittent cases, discussed further in~\sref{sec:intermittent}.

\begin{figure}[ht]
\centering
\includegraphics*[width=1\textwidth]{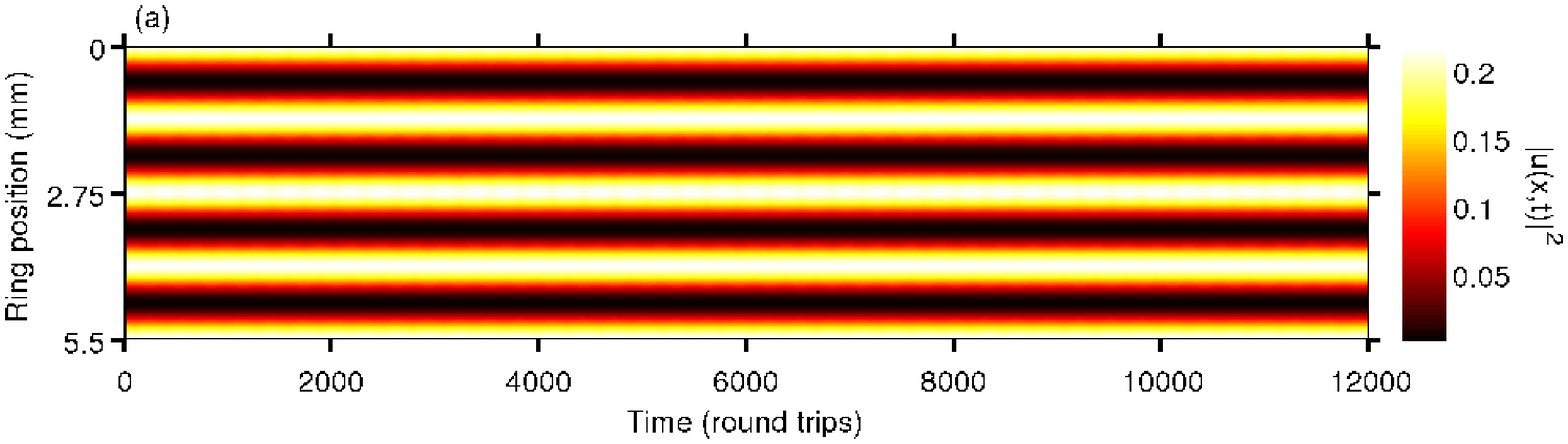}
\\
\includegraphics*[width=.48\textwidth]{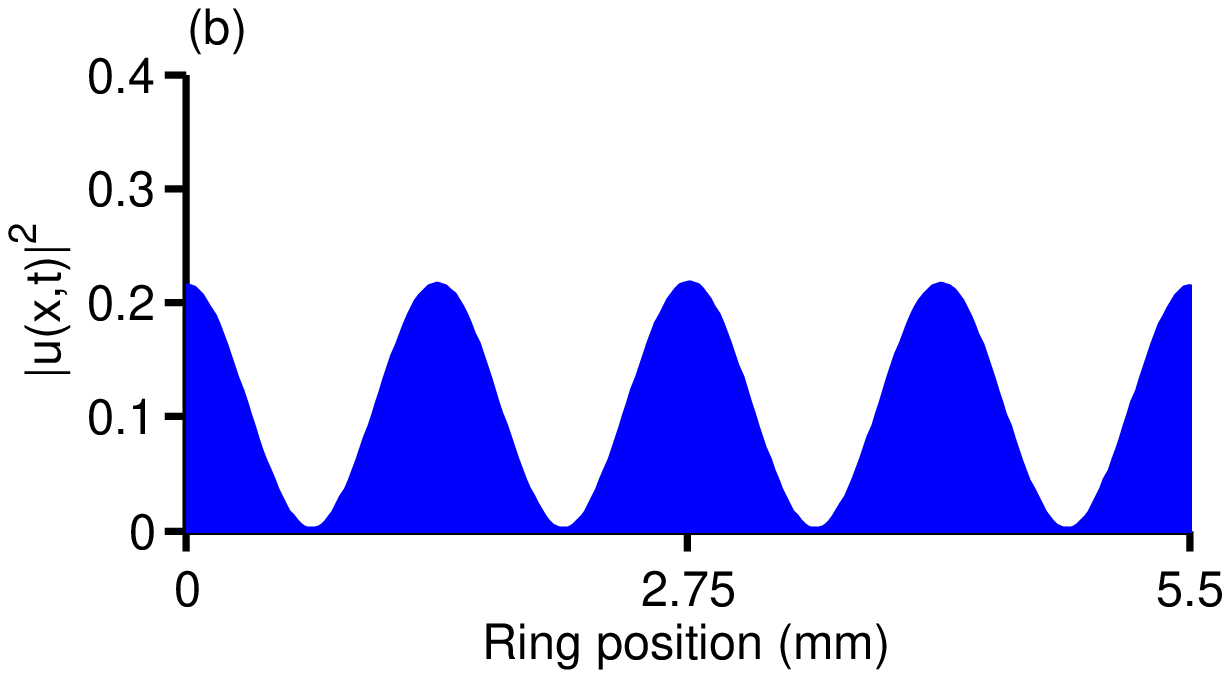}     
\includegraphics*[width=.48\textwidth]{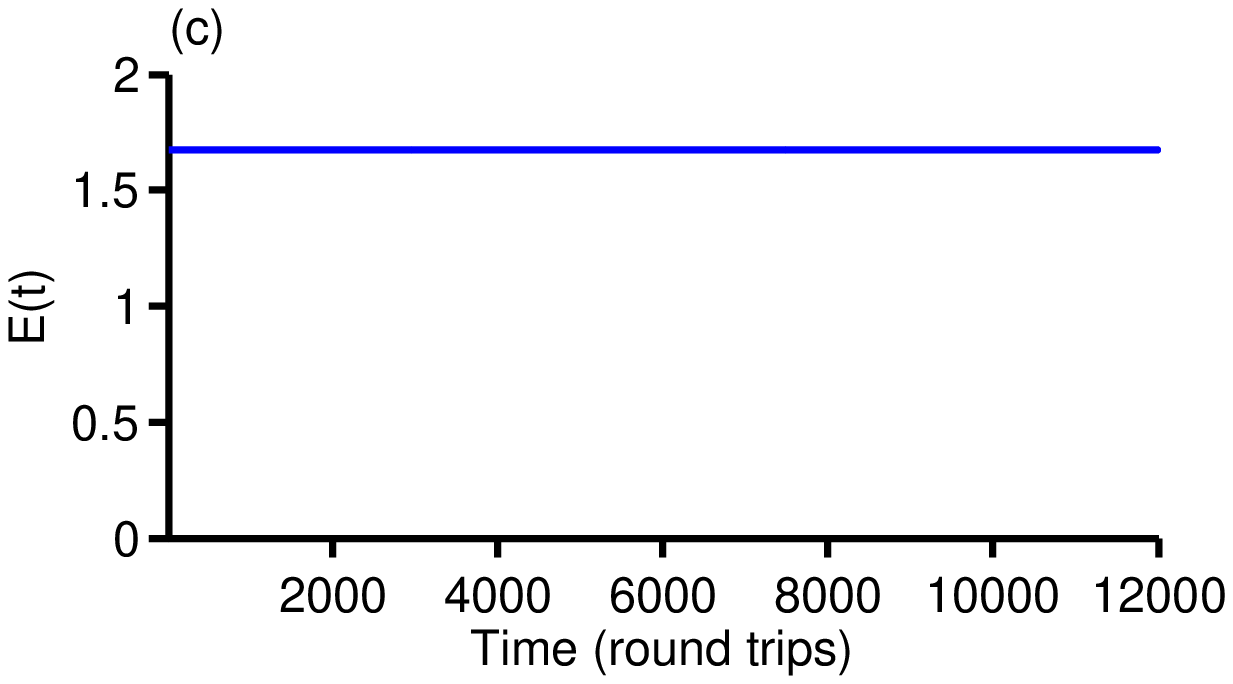}  
\caption{\label{fig:3c}Typical example of a co-propagation solution type with $N_s=4$ bright solitary waves.}
\end{figure}
\subsection{\label{sec:SS:coprop}Co-propagating Solitary Waves}

Co-propagating solitary waves are the second steady state isolated during parameter space exploration. Co-propagating solitary waves are time independent solutions where $N_s$ identical bright solitary waves propagate alongside one another without interacting. Similar $N_s$ soliton solutions of the simple cubic NLS are well studied and the number of solitons is found to be proportional to the power of the initial condition relative to the value of $N/D$~\cite{Sulem:1999}. Periodic boundary conditions require solutions have an even number of nodes. Within studied solutions $N_s$ was observed to vary from two to eight.~\Fref{fig:3c} shows the same physical quantities as plotted in~\fref{fig:3d}, with panel (c) again demonstrating the energetic stability of the solution type. The peak shape, shown in panel (b), is not consistent with either bright or dark solitons.  The example plotted in panel (a) exhibits a modulation in peak heights with a variance of $10^{-3}\%$ about the mean. While the variation is not visible in panel (a) it can be observed in an animation of the evolution, available online.

Stable co-propagation was observed only in an isolated region of parameter solutions with $L=10^{-4\pm 1}$ and $C=10^{-2}$. Quintic gains, $Q$, of orders higher than the cubic present or quintic nonlinearities, $S$, with magnitude higher than $0.01$ (the lowest order studied) drove the solution out of the steady state and in general pushed solutions into the intermittent class, discussed in~\sref{sec:intermittent}. Lower orders of quintic gain did not have any meaningful effect on stability, the number of peaks or peak height.

\section{\label{sec:intermittent}Intermittent Solutions}

Intermittent solutions demonstrate numerous distinct dynamical behaviors as the waveform evolves in time. The lifetime of these behaviors ranges from hundreds of round trips to hundreds of thousands. This corresponds to up to $1~\textrm{ms}$ before the dynamics transitions from one behavior to another. These solutions are robust to at least $10\%$ variation of initial conditions in the sense that they do not degrade to noise or experience blow-up. Such perturbations do have significant effects on the relative lifetime of each dynamical behavior and even the types of behaviors a simulation exhibits.  Quantitative matching of the intermittent dynamics to experiment will offer a challenge due to their highly transient nature; however, qualitative behaviors should be observable experimentally. In general, intermittent solutions spend a majority of their time in aperiodic evolution between distinct dynamical behaviors. Intermittent solutions can exhibit all of the behaviors previously described as temporally stable for a finite numbers of round trips.  Intermittency is the typical dynamic exhibited when stable solutions are perturbed and is therefore not observed only in isolated regions of parameter space. Stable solutions are robust to variations in initial conditions, as previously stated in~\sref{sec:meth}. Intermittency is observed when perturbations exceed $10\%$, however it bears mention that the necessary value is ultimately highly dependent on both solution type and the parameter being perturbed. Hundreds of intermittent simulations were identified during the study.

\begin{figure}[ht]
\centering
\includegraphics*[width=.65\textwidth]{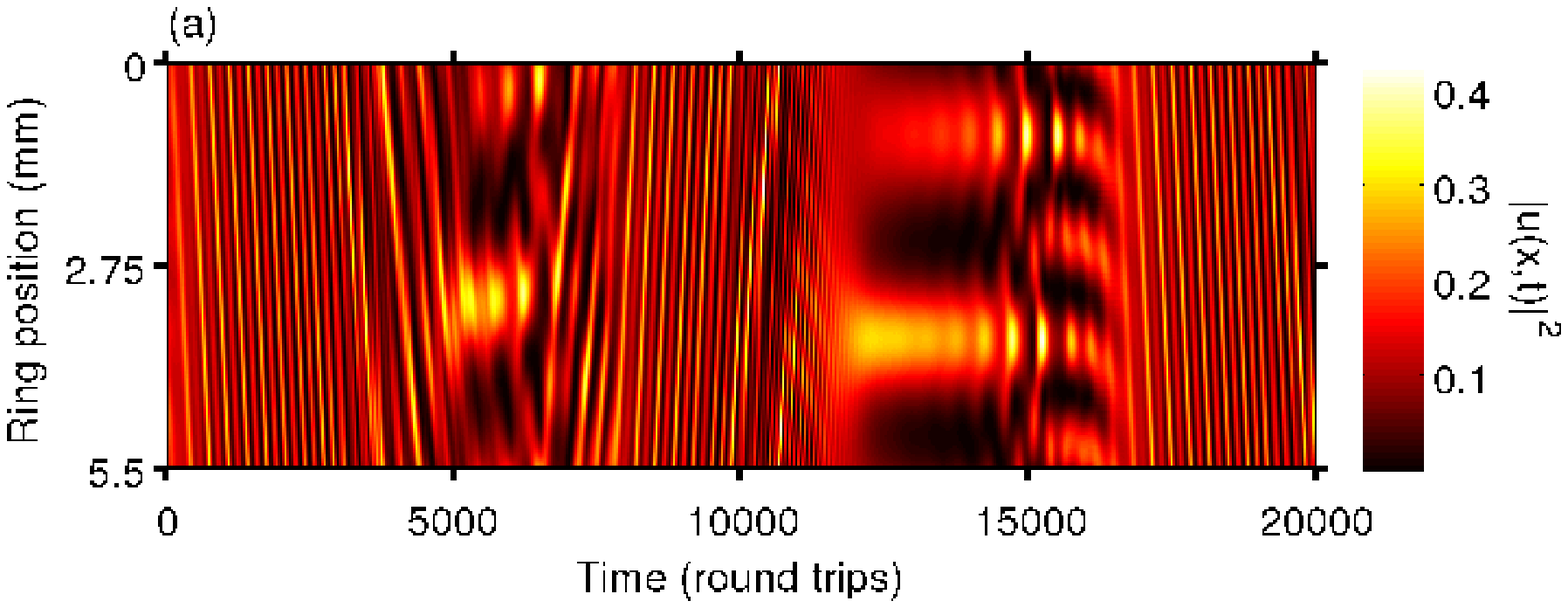}
\includegraphics*[width=.32\textwidth]{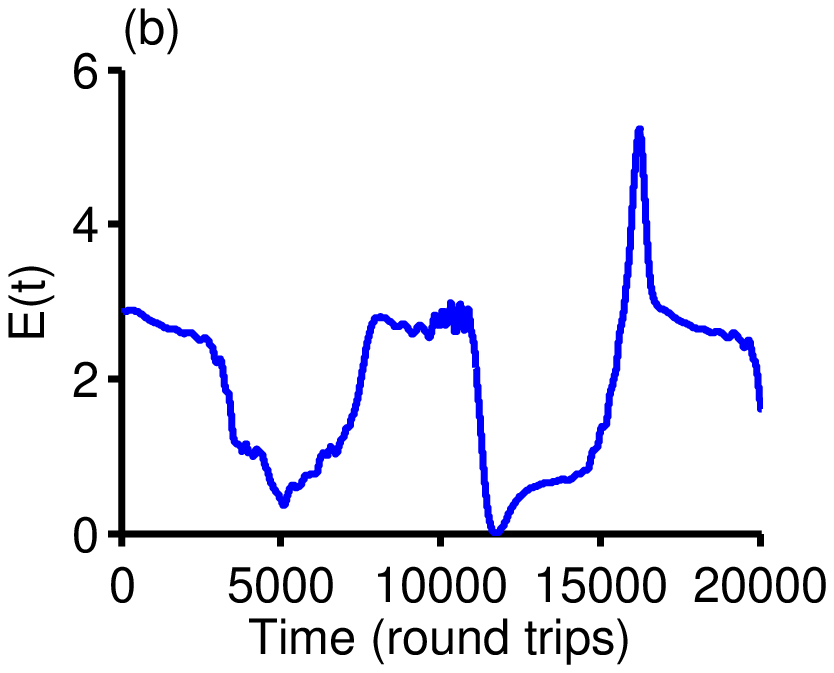}
\\
\includegraphics*[width=.65\textwidth]{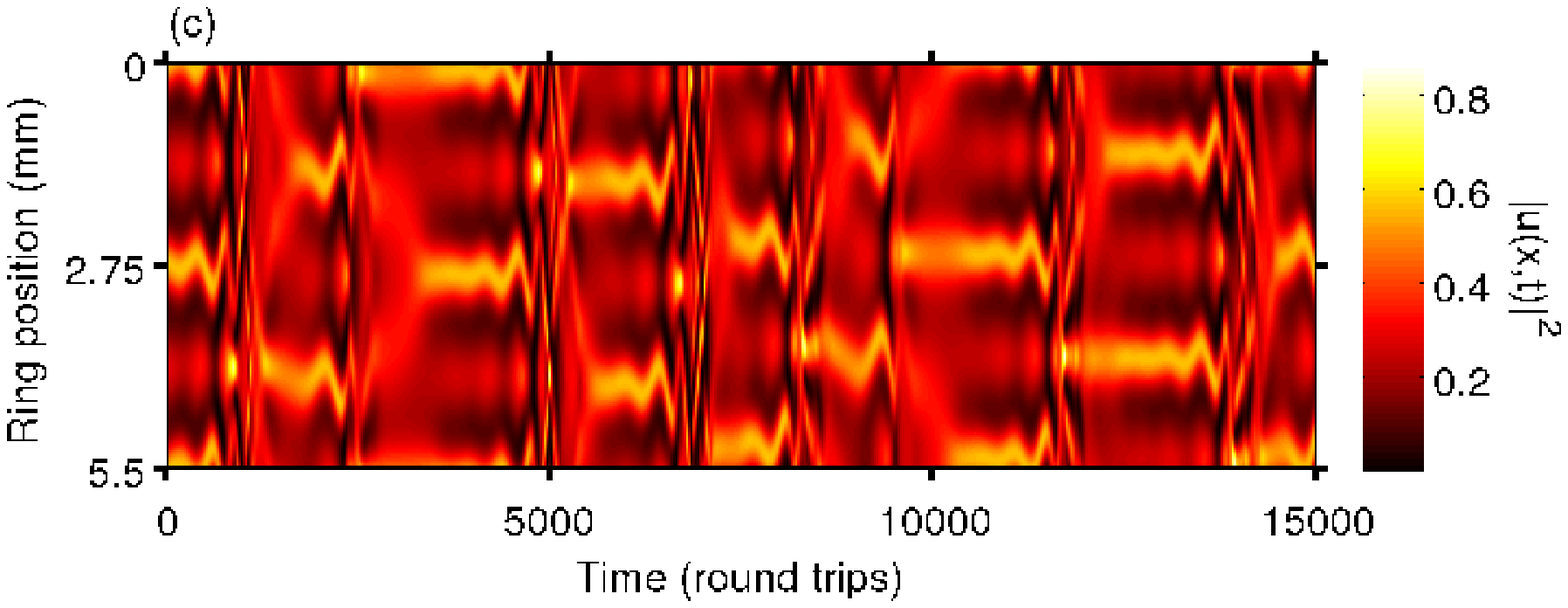}
\includegraphics*[width=.32\textwidth]{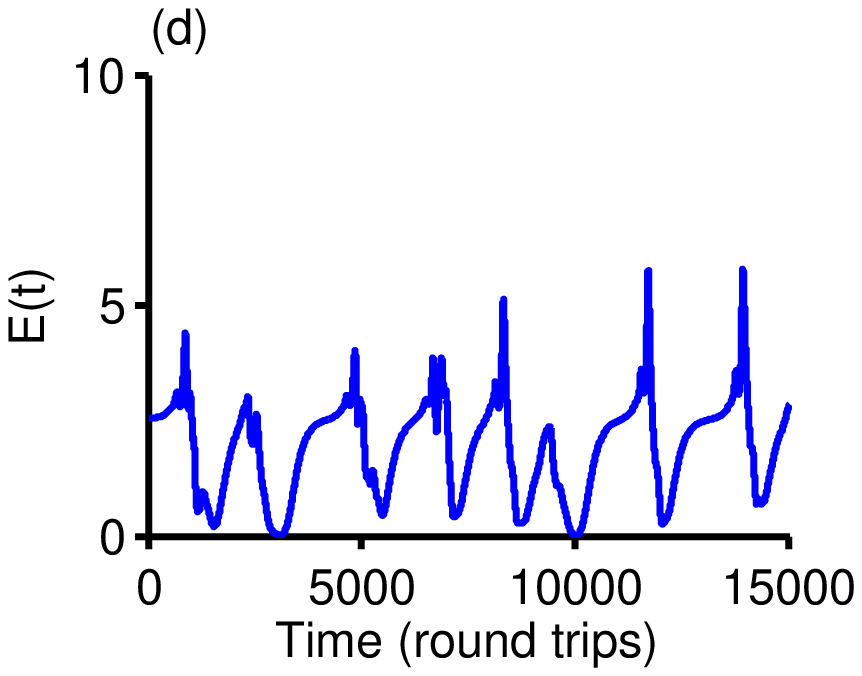}
\caption{\label{fig:int}Two examples of intermittent solutions, both exhibiting characteristic aperiodic evolution between regions of well behaved dynamics. Panel (a) shows a solution with stable regions of multi-peaked solitary waves, while panel (b) has periods of 4 peak co-propagation.}
\end{figure}

Two illustrative examples are shown in~\fref{fig:int}. The same physical quantities are shown as in~\fref{fig:EC}.  Panel (a) shows a typical simulation with three distinct multi-peaked solitary wave regimes separated by two aperiodic regimes exhibiting splitting, modulation and co-propagation behaviors. The energy is shown in plot (b) and was relatively constant during each of the multi-solitary wave regimes.  The aperiodic regimes demonstrate significantly lower energy than the finite lifetime multi-solitary wave excitations.  Panel (c) shows a simulation which exhibits periods of complex four solitary wave co-propagation interspersed with periods of aperiodic dynamics.  The lengths of successive periods of dynamical behavior are highly variable and sensitive to both parameter choice and initial condition. This sensitivity makes numerical convergence difficult to demonstrate, as discussed below in~\sref{appendix}.

\section{\label{appendix}Numerical Convergence and Quantitative vs. Qualitative Robustness}
\begin{figure}[ht]
\includegraphics*[width=.32\textwidth]{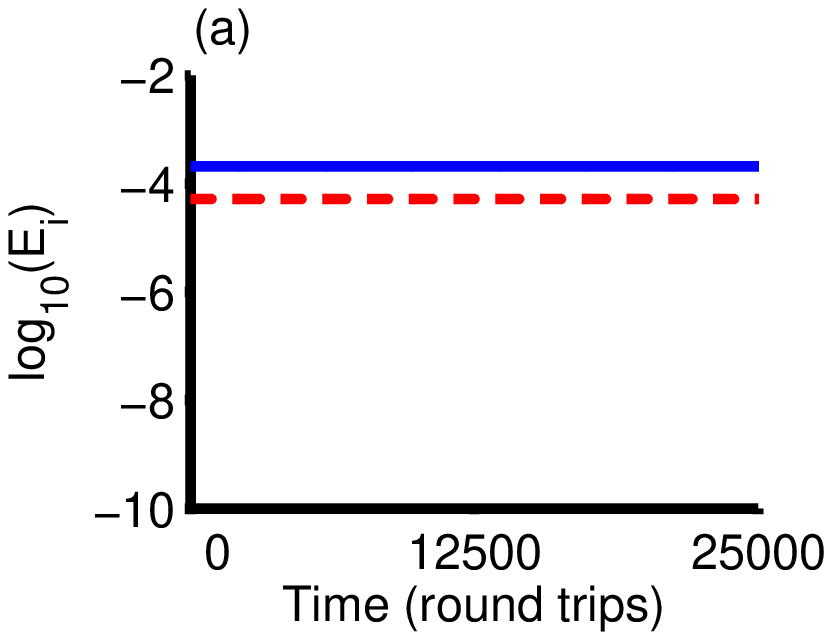}
\includegraphics*[width=.32\textwidth]{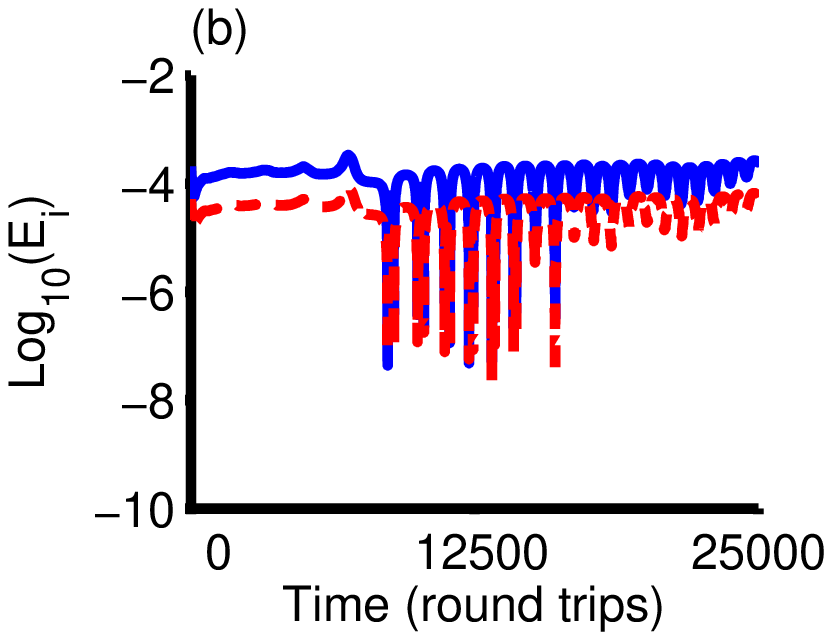}
\includegraphics*[width=.32\textwidth]{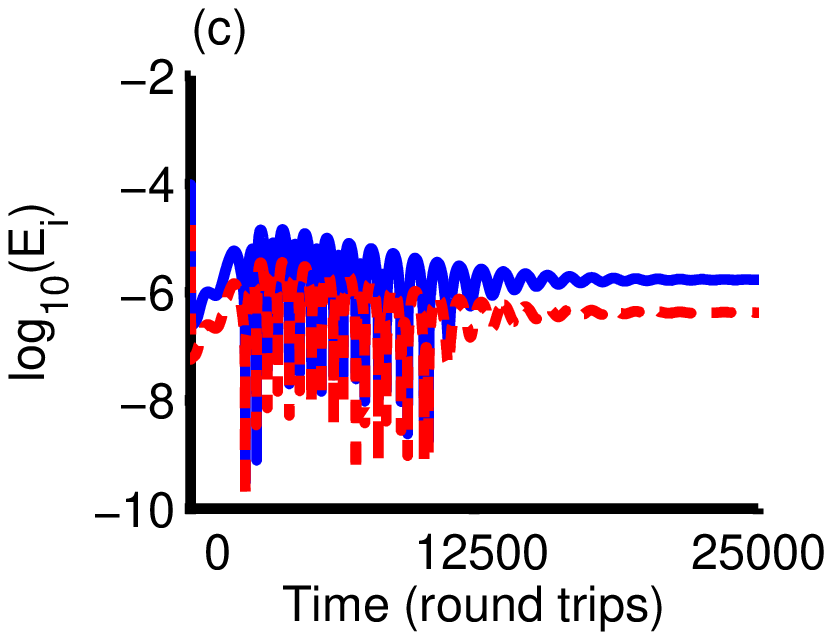}
\newline
\includegraphics*[width=.32\textwidth]{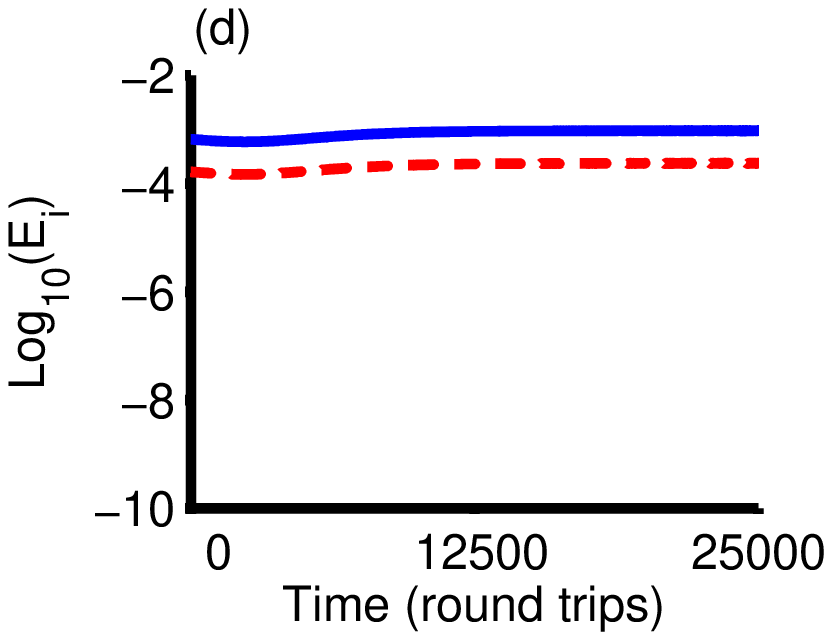}
\includegraphics*[width=.32\textwidth]{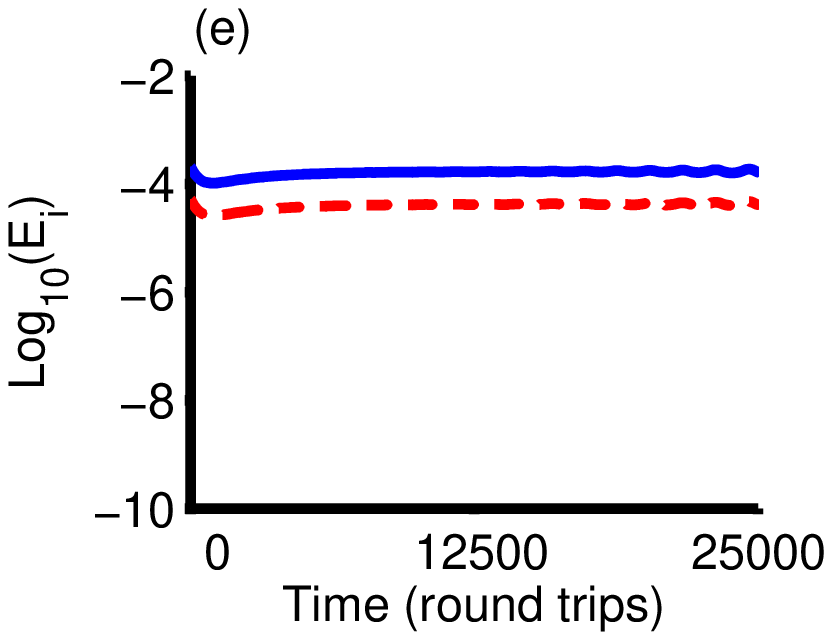}                                                                  
\caption{\label{fig:appendix:2} Relative differences in energy over 25000 round trips for increased spatial resolutions: 256 to 512 (512 to 1024) in solid blue (dashed red) line. (a) Bright soliton initial condition. (b) Solitary wave breathers.  (c) Multi-peaked soliton. (d) Co-propagation. (e) complex co-propagation.}
\end{figure}
Simulations used well established algorithms, fifth order adaptive Cash-Karp Runge-Kutta in time and pseudospectral methods in space~\cite{Snyder:2008,NR}.  Simulations were run with a spatial grid of 256 and a single step truncation error of $10^{-12}$.  The maximum number of time steps performed in a single simulation was $10^{8}$.  Initial conditions were generated using imaginary time propagation and a single step truncation error of $10^{-18}$.  Stability of initial conditions was confirmed via real time propagation.  To machine precision all initial states exhibit zero change in energy when propagated in real time for $10^9$ steps.  Initial conditions for different spatial resolutions have fixed differences in energy owning to discretization.  This discretization error decreases exponentially for increasing spatial resolution: $|E_{256}-E_{512}|=10^{-7}$, $|E_{512}-E_{1024}|=10^{-8}$ and $|E_{1024}-E_{2046}|=10^{-9}$. Results for each solution class were compared across grid sizes of 512 and 1024 and a single step truncation error of $10^{-18}$ to verify numerical convergence for algorithms used for both space and time propagation. We present convergence data for two distinct groupings of solution classes, those which exhibit extreme sensitivity to initial conditions and those which do not. The former demonstrate a qualitative robustness, while the latter are quantitatively robust.

Convergence can be demonstrated by relative difference.  Given two sets, $\left\lbrace  x_{n} \right\rbrace$ and $\left\lbrace  y_{n} \right\rbrace$, of data consisting of $N$ directly comparable observations then the relative difference at the $i^{th}$ entry is defined as
\begin{equation}\label{eqn:appendix:error}
E_{i}=\left|\frac{x_i-y_i}{x_i}\right|.
\end{equation}
This quantity offers a simple, unitless measure of the relative difference between two quantities.

\subsection{\label{appendix:2}Quantitative Robustness}
Solution classes which did not demonstrate a marked sensitivity to initial conditions were numerically converged in a traditional manner.  A distinct measurable, in this case energy, is quantitatively compared across successive time steps under different spatial and temporal resolutions.  Convergence data is graphically displayed in~\fref{fig:appendix:2} for the initial condition used during simulations,~\fref{fig:appendix:2}(a),  as well as four categorical behaviors, panels (b)-(e).  In each case the solid blue line compares spatial resolutions of 256 and 512 grid points, while the dashed red line compares the spatial resolutions of 512 and 1024 grid points.~\Fref{fig:appendix:2}(a) shows the fixed discretization error discussed previously in~\sref{appendix} while~\fref{fig:appendix:2}(b)-(e) demonstrate the spatial convergence of each dynamical behavior over the entire evolution period is as good or better than that of the initial conditions.  The greatest observed single time step relative spatial resolution error was $10^{-3}\%$.  The greatest observed single time step relative temporal resolution difference was $10^{-8}\%$. The solution types listed here were quantitatively converged:
\begin{itemize}
 \item Complex co-propagation
 \item Spatial shifting
 \item Breathers
 \item Multipeaked solitary waves
 \item Co-propagating solitary waves
\end{itemize}
\subsection{\label{appendix:1}Qualitative Robustness}
A subset of observed dynamical behaviors, from both the temporally stable and intermittent categories discussed in sections~\ref{sec:chaos}-\ref{sec:intermittent}, demonstrate an extreme sensitivity to initial conditions. These solutions were robust to variations in initial conditions and parameters of at least $10\%$ in the sense that such perturbations did not yield a shift in their categorization. However, changes in initial energy or in loss parameters of the order $10^{-9}$ and lower resulted in distinct dynamics within that categorical behavior and shifts in the starting and ending times. We note shifts in spatial resolution introduce variations of this order to the relaxed initial condition.  Therefore solutions exhibiting this sensitivity may not be converged numerically in the traditional sense. 

\begin{figure}[ht]
\includegraphics*[width=0.48\textwidth]{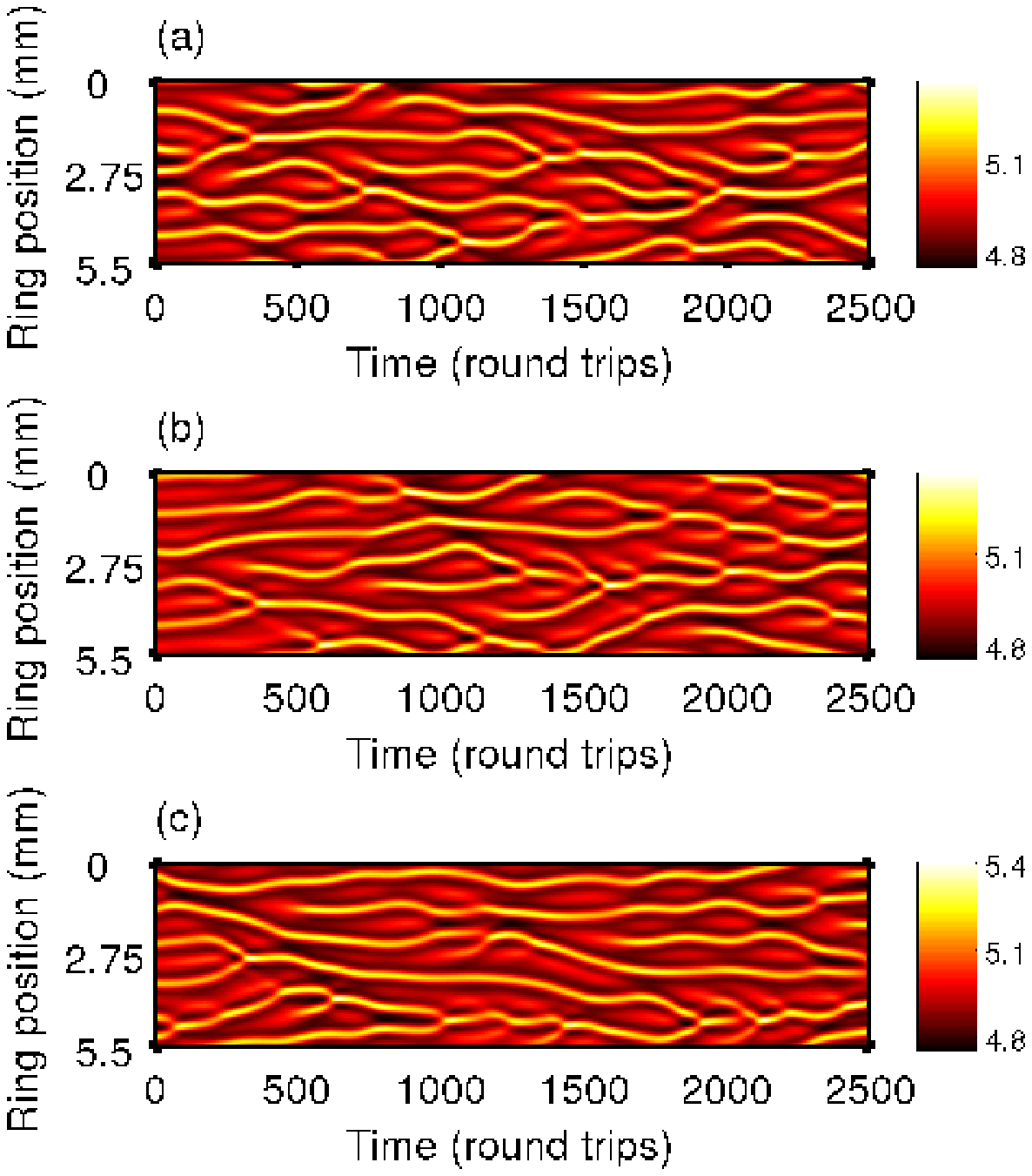}               
\includegraphics*[width=0.48\textwidth]{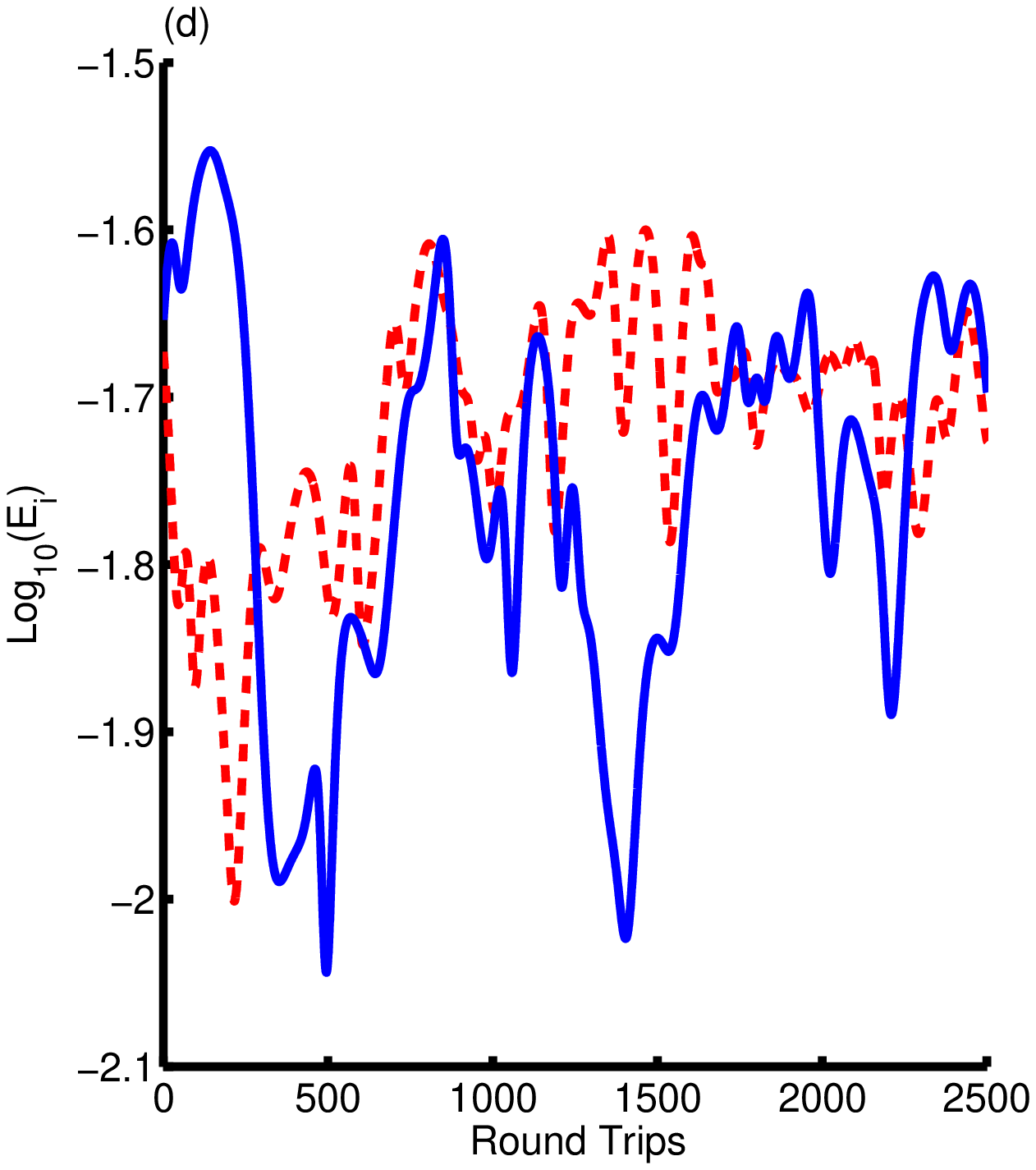}                                           
\caption{\label{fig:appendix:1}Convergence of asymmetric interaction solution type.  (a)-(c) A spatiotemporal plot of solution intensity for the same 2500 round trips for spatial resolutions of 256, 512 and 1,024 spatial grid points demonstrating stability of solution class and marked sensitivity to initial conditions. (d) Relative difference or in energy for the same round trips for 256 to 512 (512 to 1024) in solid blue (dashed red) line. Despite changes in initial conditions yielding markedly different dynamics within the solution class energies differ by less than $0.03\%$.}
\end{figure}

An illustrative example is given by the asymmetrical interaction behavior, discussed in~\sref{sec:interactions:asymmetric}, after spatial symmetry about the feedback ring center has broken.~\Fref{fig:appendix:1}(a)-(c) depicts scaled intensity across the same 2500 round trips for three different choices of spatial resolution: 256, 512 and 1024 spatial grid points respectively. The behaviors across the three spatial resolutions are qualitatively similar with each exhibiting an asymmetric $N_s$ solitary wave interaction which is characteristic of the solution type. However, the detailed dynamical behaviors of each case are quantitatively different.  Further, as shown graphically in~\Fref{fig:appendix:1}(d), the three solutions have energies which vary by less than $0.32\%$.  The behaviors listed below all demonstrated sensitivity similar to that discussed here with a relative energy difference no greater than $0.1\%$: 

\begin{itemize}
 \item Symmetric interaction
 \item Asymmetric interaction
 \item Chaotic Modulation
 \item Central Peak Recombination
\end{itemize}

This sensitivity is a typical property of evolution towards and around a strange attractor. Major attempts were made to quantify the dimensionality of the attractor as discussed in~\sref{sec:chaos} by the authors.  No stable correlation dimensions were located for any of the complex dynamical behaviors which demonstrated sensitivity to initial conditions, excluding chaotic envelope modulation. A Lyapunov exponent was estimated for the asymmetric interaction case, see~\sref{sec:interactions:asymmetric}. Further quantification and exploration of the phase space properties of GLNLS solution types and the GLNLS itself are warranted.

\subsection{\label{sec:unstable:unstable}Unstable}
Any simulation which evolved into a trivial result, degraded to noise, blew up or decayed to zero is considered to be unstable. Approximately $30\%$ of studied simulations are unstable. This is not surprising as the explored parameter space includes cases when either gain (or loss) dominate the dynamics by orders of magnitude. A subset of simulations are observed to degrade into noise due to spatial resolution issues. Pseudo-spectral methods rely on discrete fast Fourier transforms (dFFT) which provide excellent convergence for well-behaved curves. However, if the spatial features of $u(x,t)$, the complex spin wave amplitude, approach the length of the numerical lattice spacing singularities may appear and the dFFT algorithms will no longer converge locally. These errors will continue to grow and propagate over time in an ${L,Q}>0$ evolution. Further exploration of these cases with finer spatial resolutions is prohibited by computational resource constraints. In contrast, all results previously presented in sections~\ref{sec:chaos}-\ref{sec:steady_state} are converged in both space and time, as demonstrated in~\sref{appendix:2} and~\ref{appendix:1}. 

\section{\label{sec:conc}Conclusions}

We report the numerical identification of nine distinct long lifetime complex dynamical behaviors as part of six broad solution classes of the gain-loss nonlinear Schr\"odinger equation (GLNLS),~\eref{eqn:model:GLNLS}. Behaviors were located during an extensive numerical exploration of six dimensional parameter space. A minimum of eight decades were examined for each gain term while five decades of higher order nonlinearities were considered at fixed dispersion and cubic nonlinearity. The GLNLS served as a driven damped model of long lifetime spin wave dynamics in magnetic thin film active feedback rings and analogous driven damped nonlinear physical systems.  Agreement of GLNLS low dimensional chaotic modulating bright soliton trains with experimental measurements~\cite{Wang:2011} was discussed in detail.  We predicted additional GLNLS dynamical behaviors including two distinct steady state solutions, four unique examples of stable dynamical pattern formation and the intricate spatially symmetric/asymmetric interactions of solitary wave peaks. Finally we reported the existence of intermittent regimes within GLNLS parameter space, a phenomena typical of chaotic dynamical systems. Two unique examples of intermittency were presented which demonstrated finite periods of two distinct dynamical behaviors.  The variety of presented GLNLS solution types matches the scope of dynamical behaviors observed experimentally in YIG film spin wave systems, as well as predicting new behaviors that can be tested in present experiments. The GLNLS thus presents a simple yet viable and fundamental model for driven, damped nonlinear waves propagating in dispersive mediums.

We neglected the periodic effect of amplification within the feedback ring, so the gain and loss terms presented in this work represented averaged quantities.  Highly variable solution types such as the symmetric and asymmetric interactions potentially violate the GLNLS operating regime, with gain and loss driven dynamics occurring on the scale of a single round trip. Future study of this limit and adiabatically driven soliton trains is warranted. A fine grained exploration of parameter space may also be justified to identify distinct domains of stability for each observed behavior. In the future a rigorous study of GLNLS phase space would be useful to determine the cause of intermittency and potentially locate chaotic attractors of higher dimension.

This material is based upon work supported under grants number NSF PHY-0547845, NSF DMR-0906489, NSF PHY-1067973, and NSF PHY-1207881.

\section*{References}

\bibliographystyle{unsrt}
\bibliography{NJP}

\begin{thebibliography}{10}

\bibitem{Arnason:2002}
Igor~S. Aranson and Lorenz Kramer.
\newblock {The world of the complex Ginzburg-Landau equation}.
\newblock {\em Rev. Mod. Phys.}, 74:99--143, 2002.

\bibitem{Sulem:1999}
Catherine Sulem and Pierre-Louis Sulem.
\newblock {\em {The nonlinear Schr{\"o}dinger equation: {S}elf-focusing and
  wave collapse}}, volume 139.
\newblock Springer, 1999.

\bibitem{Kalinikos:1986}
B.A. Kalinikos and A.N. Slavin.
\newblock Theory of dipole-exchange spin-wave spectrum for ferromagnetic-films
  with mixed exchange boundary-conditions.
\newblock {\em J. Phys. C}, 19(35):7013--7033, 1986.

\bibitem{Slavin:1987}
A.N. Slavin and B.A. Kalinikos.
\newblock Nonlinear-theory of spin-waves in ferromagnetic films.
\newblock {\em Sov. Phys. Tech. Phys.}, 57(12):2387--2389, 1987.

\bibitem{Carusotto:2013}
Iacopo Carusotto and Cristiano Ciuti.
\newblock Quantum fluids of light.
\newblock {\em Rev. Mod. Phys.}, 85:299--366, Feb 2013.

\bibitem{Pitaevskii:2003}
Lev~Petrovich Pitaevskii and Sandro Stringari.
\newblock {\em {Bose-Einstein condensation}}.
\newblock Int. Ser. Mono. Phys. Clarendon Press, Oxford, 2003.

\bibitem{Carr:2009}
Lincoln~D Carr, David DeMille, Roman~V Krems, and Jun Ye.
\newblock Cold and ultracold molecules: science, technology and applications.
\newblock {\em New J. Phys.}, 11(5):055049, 2009.

\bibitem{Wu:2010}
Robert~E. Camley and Robert~L. Stamps, editors.
\newblock {\em Nonlinear Spin Waves in Magnetic Film Feedback Rings}, volume~62
  of {\em Sol. State Phys.}, pages 163--224.
\newblock Academic Press, 2010.

\bibitem{Slavin:1994}
A.N. Slavin and I.V. Rojdestvenski.
\newblock Bright and dark spin-wave envelope solitons in magnetic-films.
\newblock {\em IEEE Tran. Magn.}, 30(1):37--45, 1994.

\bibitem{Scott:2005}
M.M. Scott, M.P. Kostylev, B.A. Kalinikos, and C.E. Patton.
\newblock Excitation of bright and dark envelope solitons for magnetostatic
  waves with attractive nonlinearity.
\newblock {\em Phys. Rev. \rm{B}}, 71(17), 2005.

\bibitem{Kostylev:2005}
M.P. Kostylev.
\newblock Self-generation of bright spin-wave envelope solitons in active
  ferromagnetic-film rings.
\newblock {\em J. Commun. Tech. Elec.}, 50(3):313--320, 2005.

\bibitem{Kalinikos:1999}
B.A. Kalinikos, N.G. Kovshikov, and C.E. Patton.
\newblock Excitation of bright and dark microwave magnetic envelope solitons in
  a resonant ring.
\newblock {\em Appl. Phys. Lett.}, 75(2):265--267, 1999.

\bibitem{Koshikov:1996}
N.G. Kovshikov, B.A. Kalinikos, C.E. Patton, E.S. Wright, and J.M. Nash.
\newblock {Formation, propagation, reflection, and collision of microwave
  envelope solitons in Yttrium Iron Garnet films}.
\newblock {\em Phys. Rev. \rm{B}}, 54(21):15210--15223, 1996.

\bibitem{Kalinikos:2000}
B.A. Kalinikos, M.M. Scott, and C.E. Patton.
\newblock Self-generation of fundamental dark solitons in magnetic films.
\newblock {\em Phys. Rev. Lett.}, 84(20):4697--4700, 2000.

\bibitem{Kalinikos:1991}
B.A. Kalinikos, N.G. Kovshikov, and A.N. Slavin.
\newblock Envelope solitons of highly dispersive and low dispersive spin-waves
  in magnetic-films.
\newblock {\em J. Appl. Phys.}, 69(8):5712--5717, 1991.

\bibitem{Kalinikos:1990}
B.A. Kalinikos, N.G. Kovshikov, and A.N. Slavin.
\newblock {Experimental observation of magnetostatic wave envelope solitons in
  Yttrium-Iron-Garnet films}.
\newblock {\em Phys. Rev. \rm{B}}, 42(13):8658--8660, 1990.

\bibitem{Kalinikos:1990-2}
B.A. Kalinikos, N.G. Kovshikov, and A.N. Slavin.
\newblock Spin-wave envelope solitons in thin ferromagnetic-films (invited).
\newblock {\em J. Appl. Phys.}, 67(9):5633--5638, 1990.

\bibitem{Chen:1994}
M.~Chen, M.A. Tsankov, J.M. Nash, and C.E. Patton.
\newblock {Backward-volume-wave microwave-envelope solitons in
  Yttrium-Iron-Garnet films}.
\newblock {\em Phys. Rev. \rm{B}}, 49(18):12773--12790, 1994.

\bibitem{Kalinikos:1998}
B.A. Kalinikos, N.G. Kovshikov, and C.E. Patton.
\newblock Observation of self-generation of dark envelope solitons for spin
  waves in ferromagnetic films.
\newblock {\em JETP Lett.}, 68(3):243--247, 1998.

\bibitem{Kalinikos:1992}
B.A. Kalinikos, N.G. Kovshikov, and A.N. Slavin.
\newblock {Effect of magnetic dissipation on propagation of dipole spin-wave
  envelope solitons in Yttrium-Iron-Garnet films}.
\newblock {\em IEEE Trans. Magn.}, 28(5, Part 2):3207--3209, 1992.

\bibitem{Benner:2000}
H.~Benner, B.A. Kalinikos, N.G. Kovshikov, and M.P. Kostylev.
\newblock Observation of spin-wave envelope dark solitons in ferromagnetic
  films.
\newblock {\em JETP Lett.}, 72(4):213--216, 2000.

\bibitem{Slavin:2003}
A.N. Slavin and H.~Benner.
\newblock Formation and propagation of spin-wave envelope solitons in weakly
  dissipative ferrite waveguides.
\newblock {\em Phys. Rev. \rm{B}}, 67(17), 2003.

\bibitem{Wu:2004}
M.~Wu, B.A. Kalinikos, and C.E. Patton.
\newblock Generation of dark and bright spin wave envelope soliton trains
  through self-modulational instability in magnetic films.
\newblock {\em Phys. Rev. Lett.}, 93(15), 2004.

\bibitem{Kalinikos:2002}
B.A. Kalinikos, N.G. Kovshikov, M.P. Kostylev, and H.~Benner.
\newblock Self-generation of spin-wave envelope soliton trains with different
  periods.
\newblock {\em JETP Lett.}, 76(5):253--257, 2002.

\bibitem{Kalinikos:1998-2}
B.A. Kalinikos, N.G. Kovshikov, and C.E. Patton.
\newblock {Self-generation of microwave magnetic envelope soliton trains in
  Yttrium Iron Garnet thin films}.
\newblock {\em Phys. Rev. Lett.}, 80(19):4301--4304, 1998.

\bibitem{Demokritov:2003}
S.O. Demokritov, A.A. Serga, V.E. Demidov, B.~Hillebrands, M.P. Kostylev, and
  B.A. Kalinikos.
\newblock Experimental observation of symmetry-breaking nonlinear modes in an
  active ring.
\newblock {\em Nature}, 426(6963):159--162, 2003.

\bibitem{Wu:2007}
M.~Wu and C.E. Patton.
\newblock {Experimental Observation of Fermi-Pasta-Ulam Recurrence in a
  Nonlinear Feedback Ring System}.
\newblock {\em Phys. Rev. Lett.}, 98:047202, 2007.

\bibitem{Scott:2003}
M.M. Scott, B.A. Kalinikos, and C.E. Patton.
\newblock Spatial recurrence for nonlinear magnetostatic wave excitations.
\newblock {\em J. Appl. Phys.}, 94(9):5877--5880, 2003.

\bibitem{Wu:2006-2}
M.~Wu, B.A. Kalinikos, L.D. Carr, and C.E. Patton.
\newblock Observation of spin-wave soliton fractals in magnetic film active
  feedback rings.
\newblock {\em Phys. Rev. Lett.}, 96(18), 2006.

\bibitem{Wu:2006}
M.~Wu, P.~Krivosik, B.A. Kalinikos, and C.E. Patton.
\newblock Random generation of coherent solitary waves from incoherent waves.
\newblock {\em Phys. Rev. Lett.}, 96(22), 2006.

\bibitem{Wu:2005}
M.~Wu, B.A. Kalinikos, and C.E. Patton.
\newblock Self-generation of chaotic solitary spin wave pulses in magnetic film
  active feedback rings.
\newblock {\em Phys. Rev. Lett.}, 95(23), 2005.

\bibitem{Hagerstrom:2009}
Aaron~M. Hagerstrom, Wei Tong, Mingzhong Wu, Boris~A. Kalinikos, and Richard
  Eykholt.
\newblock Excitation of chaotic spin waves in magnetic film feedback rings
  through three-wave nonlinear interactions.
\newblock {\em Phys. Rev. Lett.}, 102(20), 2009.

\bibitem{Kondrashov:2008}
A.~V. Kondrashov, A.~B. Ustinov, B.~A. Kalinikos, and H.~Benner.
\newblock Chaotic microwave self-generation in active rings based on
  ferromagnetic films.
\newblock {\em Tech. Phys. Lett.}, 34(6):492--494, 2008.

\bibitem{Wu:2004-2}
M.~Wu, M.A. Kraemer, M.M. Scott, C.E. Patton, and B.A. Kalinikos.
\newblock {Spatial evolution of multipeaked microwave magnetic envelope
  solitons in Yttrium Iron Garnet thin films}.
\newblock {\em Phys. Rev. \rm{B}}, 70(5), 2004.

\bibitem{Wang:2011}
Z~Wang, A~Hagerstrom, J.Q. Anderson, W.~Tong, M.~Wu, L.D. Carr, R.~Eykholt, and
  B.A. Kalinikos.
\newblock Chaotic spin-wave solitons in magnetic film feedback rings.
\newblock {\em Phys. Rev. Lett.}, 107:114102, 2011.

\bibitem{Ustinov:2011}
Alexey~B. Ustinov, Vladislav~E. Demidov, Alexander~V. Kondrashov, Boris~A.
  Kalinikos, and Sergej~O. Demokritov.
\newblock Observation of the chaotic spin-wave soliton trains in magnetic
  films.
\newblock {\em Phys. Rev. Lett.}, 106:017201, Jan 2011.

\bibitem{Ablowitz:2001}
Mark~J. Ablowitz, G~Biondini, and S~Blair.
\newblock {Nonlinear Schr\"{o}dinger equations with mean terms in nonresonant
  multidimensional quadratic materials}.
\newblock {\em Phys. Rev. \rm{E}}, 63(4, Part 2), 2001.

\bibitem{Ablowitz:2008}
Mark~J. Ablowitz, S.A. Ablowitz, and N.~Antar.
\newblock Damping of periodic waves in physically significant wave systems.
\newblock {\em Stud. Appl. Math.}, 121(3):313--335, 2008.

\bibitem{Ablowitz:2008-2}
Mark~J. Ablowitz and Theodoros~P. Horikis.
\newblock Pulse dynamics and solitons in mode-locked lasers.
\newblock {\em Phys. Rev. \rm{A}}, 78(1), 2008.

\bibitem{Ablowitz:2008-3}
Mark~J. Ablowitz, Theodoros~P. Horikis, and Boaz Ilan.
\newblock Solitons in dispersion-managed mode-locked lasers.
\newblock {\em Phys. Rev. \rm{A}}, 77(3), 2008.

\bibitem{Akhmediev:2001}
N.~Akhmediev, J.M. Soto-Crespo, and G.~Town.
\newblock {Pulsating solitons, chaotic solitons, period doubling, and pulse
  coexistence in mode-locked lasers: complex Ginzburg-Landau equation
  approach}.
\newblock {\em Phys. Rev. \rm{E}}, 63(5):056602, 2001.

\bibitem{Akhmediev:2005}
N~Akhmediev and Adrian Ankiewicz.
\newblock Dissipative solitons in the complex ginzburg-landau and
  swift-hohenberg equations.
\newblock In {\em Dissipative solitons}, pages 1--17. Springer, 2005.

\bibitem{Akhmediev:2007}
N.~Akhmediev, J.~M. Soto-Crespo, and Ph. Grelu.
\newblock Dissipative solitons and their interactions.
\newblock {\em PAMM}, 7(1):1130301--1130302, 2007.

\bibitem{Akhmediev:2008}
N.~Akhmediev and A.~Ankiewicz.
\newblock {\em Dissipative Solitons: from Optics to Biology and Medicine},
  volume 751.
\newblock Springer, 2008.

\bibitem{Akhmediev:2009}
N.~Akhmediev, A.~Ankiewicz, J.E.M.I.A. Soto-Crespo, and P.~Grelu.
\newblock Dissipative solitons: present understanding, applications and new
  developments.
\newblock {\em Int. J. Bif. Chaos}, 19(08):2621--2636, 2009.

\bibitem{Tsoy:2005}
E.N. Tsoy and N.~Akhmediev.
\newblock {Bifurcations from stationary to pulsating solitons in the
  cubic--quintic complex Ginzburg--Landau equation}.
\newblock {\em Phys. Lett. \rm{A}}, 343(6):417--422, 2005.

\bibitem{Zhuravlev:2004}
M.N. Zhuravlev and N.V. Ostrovskaya.
\newblock {Dynamics of NLS solitons described by the cubic-quintic
  Ginzburg-Landau equation}.
\newblock {\em J. Exp. Theo. Phys.}, 99(2):427--442, 2004.

\bibitem{Stancil:2009}
D.D. Stancil and A.~Prabhakar.
\newblock {\em Spin Waves: Theory and Applications}.
\newblock Springer, 2009.

\bibitem{Krivosik:2012}
Pavol Krivosik and Carl~E. Patton.
\newblock {Hamiltonian formulation of nonlinear spin-wave dynamics: Theory and
  applications}.
\newblock {\em Phys. Rev. \rm{B}}, 82:184428, 2010.

\bibitem{Leblond:2001}
H.~Leblond.
\newblock {Rigorous derivation of the NLS in magnetic films}.
\newblock {\em J. Phys. \rm{A}}, 34(45):9687, 2001.

\bibitem{Snyder:2008}
Victor Snyder.
\newblock {\em {Quantum Fluctutations in the Dynamics of Bose-Einstein
  Condensates}}.
\newblock PhD thesis, Colorado School of Mines, 2008.

\bibitem{NR}
W.H. Press, S.A. Teukolsky, W.T. Vetterling, and B.P. Flannery.
\newblock {\em Numerical recipes 3rd edition: The art of scientific computing}.
\newblock Cambridge University Press, 2007.

\bibitem{Kosloff:1986}
R~Kosloff and H~Tal-Ezer.
\newblock {A direct relaxation method for calculating eigenfunctions and
  eigenvalues of the Schr{\"o}dinger equation on a grid}.
\newblock {\em Chem. Phys. Lett.}, 127(3):223--230, 1986.

\bibitem{Scott:2004}
M.M. Scott, C.E. Patton, M.P. Kostylev, and B.A. Kalinikos.
\newblock {Nonlinear damping of high-power magnetostatic waves in
  Yttrium-Iron-Garnet films}.
\newblock {\em J. Appl. Phys.}, 95(11, Part 1):6294--6301, JUN 2004.

\bibitem{Hegger:1999}
R.~Hegger, H.~Kantz, and T.~Schreiber.
\newblock {Practical implementation of nonlinear time series methods: The
  TISEAN package}.
\newblock {\em Chaos}, 9(2):413--435, 1999.

\bibitem{Kantz:2004}
H.~Kantz and T.~Schreiber.
\newblock {\em Nonlinear time series analysis}.
\newblock Cambridge Univ Pr, 2004.

\bibitem{Takens:1981}
Floris Takens.
\newblock Detecting strange attractors in turbulence.
\newblock In {\em Dynamical systems and turbulence}, pages 366--381. Springer,
  1981.

\bibitem{Sauer:1991}
T.~Sauer, J.A. Yorke, and M.~Casdagli.
\newblock Embedology.
\newblock {\em J. Stat. Phys.}, 65(3):579--616, 1991.

\bibitem{Theiler:1990}
J.~Theiler.
\newblock Estimating fractal dimension.
\newblock {\em J. Opt. Soc. Am. \rm{A}}, 7(6):1055--1073, 1990.

\bibitem{Cross:1993}
Mark~C Cross and Pierre~C Hohenberg.
\newblock Pattern formation outside of equilibrium.
\newblock {\em Rev. Mod. Phys,}, 65(3):851, 1993.

\end{thebibliography}

\end{document}